\documentclass[a4paper,11pt]{article}
\usepackage{booktabs, jheppub, slashed} 
\usepackage[utf8x]{inputenc} 
\usepackage{comment,diagbox}

\hypersetup{colorlinks=true, linkcolor=red, urlcolor=blue, citecolor=red}

\preprint{CTPU-PTC-23-43}

\title{A novel search strategy for right-handed charged gauge bosons at the Large Hadron Collider}

\author[a]{Mariana Frank}
\author[b]{\!\!, Benjamin Fuks}
\author[c]{\!\!, Adil Jueid}
\author[d,e]{\!\!, Stefano Moretti}
\author[a,d]{\! and \"{O}zer \"{O}zdal}

\affiliation[a]{Department of Physics, Concordia University, 7141 Sherbrooke St. West, Montreal, Quebec H4B 1R6, Canada}
\affiliation[b]{ Laboratoire de Physique Th\'{e}orique et Hautes \'{E}nergies (LPTHE), UMR 7589,\\ Sorbonne Universit\'{e} \& CNRS, 4 place Jussieu, 75252 Paris Cedex 05, France}
\affiliation[c]{Particle Theory and Cosmology Group, Center for Theoretical Physics of the Universe, Institute for Basic Science (IBS), Daejeon, 34126, Republic of Korea}
\affiliation[d]{School of Physics $\&$ Astronomy, University of Southampton, Highfield, Southampton SO17 1BJ, UK}
\affiliation[e]{Department of Physics $\&$  Astronomy, Uppsala University, Box 516, 75120 Uppsala, Sweden}

\emailAdd{mariana.frank@concordia.ca}
\emailAdd{fuks@lpthe.jussieu.fr}
\emailAdd{adiljueid@ibs.re.kr}
\emailAdd{s.moretti@soton.ac.uk; stefano.moretti@physics.uu.se}
\emailAdd{ozer.ozdal@soton.ac.uk}

\begin{document}

\abstract{
  We explore the potential of the Large Hadron Collider (LHC) in detecting a signal originating from the production of a heavy $SU(2)_R$ charged gauge boson that then decays into a top-bottom quark pair via the mediation of a right-handed neutrino, $p p \to W_R \to N_R \ell \to (\ell' t b)\ell$. Such a channel, that we study in the context of the minimal Left-Right Symmetric Model, contrasts with conventional smoking-gun signatures targeted experimentally and phenomenologically  in which only light quarks are involved. We propose a selection strategy aimed at extracting such a top-bottom signal and we estimate the resulting sensitivity of the LHC to the model. Our results demonstrate the potential impact of such a search and we therefore urge the experimental collaborations to carry out a similar analysis in the light of present and future data.
}

\maketitle

%%%%%%%%%%%%%%%%%%%%%%%%%%%%%%%%
\section{Introduction}
\label{sec:intro}
%%%%%%%%%%%%%%%%%%%%%%%%%%%%%%%%

The Standard Model (SM) of particle physics is unable to address fundamental questions in Nature despite being in excellent agreement with all measurements carried out at experiments so far. For instance, the origin of parity violation in electroweak (EW) interactions remains a mystery to which the SM has no answer. Left-Right Symmetric Models (LRSMs) are one of the most attractive solutions to such a problem  \cite{Pati:1974yy, Mohapatra:1974hk, Senjanovic:1975rk, Senjanovic:1978ev}. Their minimal incarnation is dubbed the minimal LRSM (mLRSM). It is based on the $SU(2)_L \otimes SU(2)_R \otimes U(1)_{B-L}$ gauge group, and it predicts a particle spectrum comprising several extra states, including new charged and neutral gauge bosons, extra (pseudo)scalars (\textit{i.e.}~neutral, singly-charged and doubly-charged bosons), and three generations of right-handed neutrinos. Furthermore, this minimal model accommodates naturally small neutrino masses through the seesaw mechanism~\cite{Mohapatra:1979ia, Mohapatra:1980yp, Cai:2017mow}, it elegantly addresses the problem of parity violation, and it has a rich phenomenology both at colliders and relative to low-energy experiments. The presence of extra right-handed charged-currents and new (pseudo)scalar interactions indeed impact lepton flavour violation in charged lepton decays and $e \to \mu$ conversions~\cite{Cirigliano:2004mv}, EW precision observables~\cite{Hsieh:2010zr, Blanke:2011ry, Bernard:2020cyi}, Charge-Parity (CP) violations in meson decays and meson-antimeson oscillations~\cite{Beall:1981ze, Mohapatra:1983ae, Ecker:1983uh, Ecker:1985ei, Frere:1991db, Ball:1999mb, Maiezza:2010ic, Bertolini:2014sua}, electric dipole moments of various particle types~\cite{Ecker:1983dj, Frere:1991jt, Xu:2009nt, Cirigliano:2016yhc, Ramsey-Musolf:2020ndm}, and nuclear $\beta$ decays~\cite{Cirigliano:2013xha, Gonzalez-Alonso:2018omy}. In particular, a recent comprehensive global analysis of the mLRSM by means of low-energy observables has been carried in Ref.~\cite{Dekens:2021bro}, while the consequences on its leptonic and  (pseudo)scalar sectors have been analysed in Refs.~\cite{Prezeau:2003xn, Tello:2010am, Nemevsek:2011aa, Nemevsek:2012iq, Barry:2013xxa, BhupalDev:2014qbx, Bambhaniya:2015ipg} and \cite{Barenboim:2001vu, Maiezza:2016ybz, Bambhaniya:2013wza, Dev:2016dja} respectively. In contrast to other models featuring multi-Higgs doublets, the extra (pseudo)scalar states of the mLRSM originate from an $SU(2)_L \otimes SU(2)_R$ bi-doublet, and it has been shown that they need to be as heavy as ${\cal O}(10)$ TeV to avoid large flavour-changing neutral-current transitions. 

The mLRSM also predicts lepton number violation that contribute to neutrinoless double-beta ($0\nu\beta\beta$) decay rates. Such contributions are highly correlated with the neutrino mass parameters and the $SU(2)_R$ gauge boson masses, while being amenable to discovery even for small neutrino masses (see Refs.~\cite{Rodejohann:2011mu, Cirigliano:2018yza, Li:2020flq, Yang:2021ueh} for recent analyses).  The mLRSM correspondingly features high-energy analogues of the usual low-energy $0\nu\beta\beta$ processes. One of them consists of the so-called Keung-Senjanovic mechanism that is expected to occur at high-energy hadron colliders~\cite{Keung:1983uu}. The idea is that on-shell and off-shell $SU(2)_R$ gauge bosons, denoted in what follows as $W_R$, can decay into a lepton and right-handed neutrino $N_R$ that itself further decays into two light quarks and another (charged) lepton. If the two charged leptons have the same electric charge, then this process breaks lepton number conservation by two units, providing therefore a direct correlation with the rates of $0\nu\beta\beta$ decays. Furthermore, the production of opposite-sign charged leptons is also interesting on its own~\cite{Nemevsek:2011hz, Das:2017hmg, Nemevsek:2018bbt, Nemevsek:2023hwx}, which has consequently motivated the ATLAS and CMS collaborations to carry out several searches for a $W_R$ state at the Large Hadron Collider (LHC), both in the same-sign and opposite-sign dileptonic channels~\cite{ATLAS:2018dcj, ATLAS:2019isd, CMS:2018agk, CMS:2018jxx, CMS:2021dzb}. Depending on the $N_R$ mass and the lepton flavour structure, $W_R$-boson masses up to $4.8$--$5.0$~TeV are excluded today~\cite{CMS:2021dzb}.\footnote{These bounds can be relaxed in cases where the $W_R$ boson couples in a generic way to leptons and quarks, {\it i.e.}~when $g_L \neq g_R$ (see \textit{e.g.}~Refs. \cite{Frank:2018ifw, Ozdal:2021rlj}).} Furthermore, at large momentum transfer the same $0\nu\beta\beta$ matrix elements imply several enhancements in the production rates of same-sign lepton pairs of possibly different flavours via vector-boson scattering~\cite{Fuks:2020att, Fuks:2020zbm, CMS:2022hvh, ATLAS:2023tkz}. Constraints on heavy neutrinos with masses in the 50~GeV -- 20~TeV range can here be imposed, hence even above the LHC energy scale.

In the light of these efforts, an interesting question, that was also raised in Ref.~\cite{Mitra:2016kov}, arises: {\sl what if we  use heavy quarks instead of light ones in searching for $W_R$ and $N_R$ states?} In this case the produced $N_R$ particle undergoes a three-body decay into a charged lepton, a bottom quark,  and a top quark, leading thus to a very rich final state. This consists of an intriguing signature, as it is widely believed that the top quark can play an important role in probing new physics Beyond the SM (BSM). % (see for example Refs.~\cite{Agram:2013wda, Arina:2016cqj, Arhrib:2018bxc, Godbole:2019erb, Chatterjee:2019brg, Cheung:2020ugr, Bhaskar:2020gkk, Darme:2021gtt} for analyses related to various physics cases).

The aim of this study is to propose a {\sl novel} search strategy for the production of an $SU(2)_R$ charged gauge boson $W_R$ and a  neutrino $N_R$ at the LHC, relying on the $W_R\to N_R \ell\to t{b} \ell \ell$ decay chain with the $W_R$ emerging from charged current Drell-Yan (DY) production. The top quark, being produced from a heavy neutrino decay, is highly boosted in most of the cases. Jet substructure methods and top taggers have therefore the potential to efficiently reduce the SM background, together with specific kinematics variables exploiting the richness of the final state. The use of this channel can thus be crucial not only for discovery purposes but also for diagnostics as a probe of the properties of the mLRSM as a whole. 

The rest of this paper is organised as follows. In section \ref{sec:theo} we briefly discuss the mLRSM, its field content and the associated Lagrangian, and we assess the constraints emerging from recent searches for $W_R$ bosons and $N_R$ neutrinos at colliders (with some technical details on the reinterpretation of the LHC results being reported in Appendix~\ref{sec:MA5:CMS}). Next we study $W_R$ and $N_R$ production and decay in the $\ell\ell tb$ channel in section \ref{sec:tbll}, where we also present our signal and background analysis, and its results. The latter include an estimation of the sensitivity of the LHC to the mLRSM signal considered, and we demonstrate the potential usefulness of designing corresponding searches in real LHC data. We conclude and summarise our work in section \ref{sec:conclusions}.
%%%%%%%%%%%%%%%%%%%%
\section{The model and current bounds}
\label{sec:theo}
%%%%%%%%%%%%%%%%%%%%

%%%%%%%%%%%%%%%%%%%%%%%%%%%%%%%%%%%%%%%%%%%%%%%
\subsection{The model}
%%%%%%%%%%%%%%%%%%%%%%%%%%%%%%%%%%%%%%%%%%%%%%%%

In this section, we briefly describe the mLRSM, its particle content, and the interaction Lagrangian relevant for our analysis (more technical details can be found in Refs.~\cite{Tello:2012qda, Chakrabortty:2013mha, Maiezza:2016bzp, BhupalDev:2018xya, Chauhan:2019fji}). The fermion sector of the mLRSM includes the following fields:
\begin{equation}
  q_L \equiv \begin{pmatrix} u_L \\ d_L \end{pmatrix}_{({\bf 2}, {\bf 1}, \frac13)},\ \ 
  q_R \equiv \begin{pmatrix} u_R  \\ d_R \end{pmatrix}_{({\bf 1}, {\bf 2}, \frac13)}, \ \
  \ell_L \equiv \begin{pmatrix} \nu_L \\  e_L \end{pmatrix}_{({\bf 2}, {\bf 1}, -1)}, \ \
  \ell_R \equiv \begin{pmatrix} N_R \\  e_R\end{pmatrix}_{({\bf 1}, {\bf 2}, -1)},
\label{eq:fermions}
\end{equation}
in which the subscripts refer to the representation of the fermion fields under $SU(2)_L \otimes SU(2)_R \otimes U(1)_{B-L}$. In addition, we denote the three gauge couplings associated with this gauge symmetry by $g_L$, $g_R$, and $g_{B-L}$. Parity conservation at high scales dictates that the gauge interactions be invariant under 
\begin{eqnarray}
    \{W_L, q_L, \ell_L \} \longleftrightarrow \{W_R, q_R, \ell_R\}.
\end{eqnarray}
An immediate consequence of this symmetry is that $g_L$ and $g_R$ are equal, {\it i.e.}~$g_L \equiv g_R \equiv g$. Thus EW symmetry breaking is minimally achieved through the vacuum expectation values acquired by the neutral components of three scalar multiplets: a bi-doublet ($\Phi$) and two triplets ($\Delta_{L,R}$) represented as:
\begin{equation}\renewcommand{\arraystretch}{1.3}\setlength{\arraycolsep}{6pt}
    \Phi \equiv \begin{pmatrix} \phi_1^0  &  \phi_2^+ \\ \phi_1^- & \phi_2^0 \end{pmatrix}_{({\bf 2}, {\bf 2}, 0)},\ \
    \Delta_L \equiv \begin{pmatrix}  \frac{\delta_L^+}{\sqrt{2}}   & \delta_L^{++}  \\ \delta_L^0  &   -\frac{\delta_L^+}{\sqrt{2}} \end{pmatrix}_{({\bf 3}, {\bf 1}, 2)},\ \
    \Delta_R \equiv \begin{pmatrix}  \frac{\delta_R^+}{\sqrt{2}}   & \delta_R^{++}  \\ \delta_R^0  &   -\frac{\delta_R^+}{\sqrt{2}} \end{pmatrix}_{({\bf 1}, {\bf 3}, 2)}.
\end{equation}
The model's Lagrangian $\mathcal{L}$ is given by
\begin{equation}\label{eq:lag}\begin{split}
    & \mathcal{L} = {\cal L}_{\rm gauge} + i \bar{q}_L \slashed{D} q_L + i \bar{q}_R \slashed{D} q_R  + i \bar{\ell}_L \slashed{D} \ell_L + i \bar{\ell}_R \slashed{D} \ell_R \\
    & \ + {\rm Tr}\Big[(D_\mu \Phi)^\dagger (D^\mu \Phi) \Big] + {\rm Tr}\Big[(D_\mu \Delta_L)^\dagger (D^\mu \Delta_L) \Big] + {\rm Tr}\Big[(D_\mu \Delta_R)^\dagger (D^\mu \Delta_R) \Big] \\
    & \ - \bigg[\bar{q}_L \big(Y_q \Phi \!+\! \tilde{Y}_q \tilde{\Phi}\big) q_R + \bar{\ell}_L \big(Y_\ell \Phi \!+\! \tilde{Y}_\ell  \tilde{\Phi}\big) \ell_R  + \bar{\ell}_L^c i \sigma_2 \Delta_L Y_L \ell_L + \bar{\ell}_R^c i \sigma_2 \Delta_R Y_R \ell_R + {\rm H.c.} \bigg] \\
    & \ - V(\Phi, \Delta_L, \Delta_R),
\end{split}\end{equation}
where ${\cal L}_{\rm gauge}$ includes kinetic terms for all gauge bosons, $V(\Phi, \Delta_L, \Delta_R)$ is the scalar potential (which exact form is irrelevant for our study), and all Yukawa couplings $Y$ are $3\times3$ matrices in the flavour space. Moreover, $\tilde\Phi$ is the dual bi-doublet Higgs field, and $D_\mu$ stands for the covariant derivative operator
\begin{eqnarray}
    D_\mu = \partial_\mu - i g T_L^I W^I_{L,\mu} - i g T_R^I W^I_{R,\mu} - i \frac{g_{B-L}}{2}Q_{B-L} B_\mu,
\end{eqnarray}
with $T_L^I$ and $T_R^I$ being the generators of $SU(2)_L$ and $SU(2)_R$ taken in the relevant representation, and $Q_{B-L}$ is the associated $U(1)_{B-L}$ charge. EW symmetry breaking proceeds in two steps. First, the vacuum expectation value acquired by the triplet $\Delta_R$ breaks the $SU(2)_R \otimes U(1)_{B-L}$ symmetry down to the hypercharge group $U(1)_Y$. Second, the bi-doublet $\Phi$ and triplet $\Delta_L$ Higgs fields choose a configuration breaking the $SU(2)_L \otimes U(1)_Y$ symmetry down to electromagnetism. The full corresponding vacuum configuration is hence given by
\begin{eqnarray}
   \langle \Phi \rangle = \left(\begin{array}{cc}
       v_1  &  0 \\
       0 & -v_2 e^{-i\alpha} 
    \end{array}\right), \qquad 
   \langle \Delta_{L,R} \rangle \equiv \left(\begin{array}{cc}
      0   & 0  \\
      v_{L,R}  &  0 
    \end{array}\right).
\end{eqnarray}
We further define the mixing angle $\beta$ such that $v = v_1 \sin\beta =v_2 \cos\beta$, and we enforce the hierarchy $v_L \ll v \ll v_R$ to get  agreement with neutrino data ($v_L$ being small) and constraints on additional gauge bosons ($v_R$ being large). In this setup, the masses of the charged gauge bosons read
\begin{eqnarray}
    M_{W_L}^2 \approx \frac{1}{2} g^2 v^2,  \qquad M_{W_R}^2 \approx g^2 v_R^2.
\end{eqnarray}

%%%%%%%%%%%%%%%%%%%%%%%%%%%%%%%%%%%%%%%%%%%%%%%%%%%%%%%%%%%%%%%%%%%%%%%%%%%%%%%%%%%%%%%
\subsection{Limits on $W_R$ properties from collider and low-energy experiments}
\label{sec:bounds}
%%%%%%%%%%%%%%%%%%%%%%%%%%%%%%%%%%%%%%%%%%%%%%%%%%%%%%%%%%%%%%%%%%%%%%%%%%%%%%%%%%%%%%%
Several searches have focused on testing for the existence of $W_R$ bosons associated with the $SU(2)_L \times SU(2)_R \times U(1)_{B-L}$ gauge symmetry. While at tree-level the mixing between the two electrically charged bosons $W_L$ and $W_R$ results in a shift of the $W_L$ mass from its SM value $M_W$, data indicate that the corresponding mixing angle must be smaller than $10^{-2}$~\cite{ParticleDataGroup:2022pth}. This is consistent with our model configuration in which this mixing is negligible, and in which the $W_R$ boson couples essentially to right-handed fermions. The structure of such a coupling in the flavour space is further dictated by the values of the elements of a Cabibbo-Kobayashi-Maskawa (CKM) matrix relating right-handed fermions, which generally needs not be proportional to the known CKM matrix (that relates left-handed fields).

At hadron colliders searches for signatures of a $W_R$ boson have concentrated on resonant production. The most commonly looked for signal is made up of high-momentum electrons or muons accompanied by a large amount of missing transverse energy, originating from the process
\begin{equation}\label{eq:WRDY}
  pp \to W_R X \to \ell N_R X\qquad\text{with}\qquad\ell = e, \mu, \tau.
\end{equation}
Searches in this channel assume that the narrow width approximation is valid (with the $W_R$ boson width-over-mass ratio $\Gamma_{W_R}/M_{W_R} \leq 7\%$)\footnote{The effects of the $W_R$-boson width on the signal have been explored in Refs.~\cite{Accomando:2011eu, Accomando:2013slb}.}, and that the heavy neutrino $N_R$ is lighter than the $W_R$ boson and escapes detection. Relying on the Sequential Standard Model (SSM) as a benchmark new physics setup, the ATLAS and CMS collaborations have set stringent limits on the $W_R$ boson by making use of 139~fb$^{-1}$ of data at a centre-of-mass energy $\sqrt{s}=13$ TeV. In the electronic channel, they constrained the $W_R$ boson mass to satisfy $M_{W_R} > 6$~TeV, while in the muonic channel the bounds are reduced to 5.6~TeV~\cite{ATLAS:2019lsy, CMS:2022krd}. % Comparatively weaker limits exist for $M_{W_R} \le 150$ GeV from Tevatron data at $\sqrt{s}=1.8$ TeV~\cite{CDF:1995udz}. 
In contrast, limits only reach $M_{W_R}>5$ TeV for a final state comprising tau leptons. By virtue of the Keung-Senjanovic mechanism, the right-handed neutrino $N_R$ produced through  process~\eqref{eq:WRDY} could also decay, through a virtual $W_R$ boson exchange, into an $ee jj$, $\mu \mu jj$ or $\tau\tau jj$ system. Such a new physics signature was explored by both collaborations, and cross section limits as functions of the $N_R$ and $W_R$ masses have been determined from 13~TeV LHC data~\cite{CMS:2018iye, CMS:2021dzb, ATLAS:2019isd}. This is further addressed in section~\ref{sec:cmsbounds}.

Searches for di-jet resonances can also be used to set bounds on the signal originating from the process $pp\to W_R \to q {\bar q}^\prime$. Within the SSM, limits on the $W_R$ boson mass of 4~TeV have been obtained~\cite{CMS:2021dzb, ATLAS:2019fgd}. On the other hand, The ATLAS collaboration also excluded $M_{W_R}$ lighter than $3.25$ TeV from $p p \to W_R \to t \bar{b}+{\rm h.c.}$ using $36~{\rm fb}^{-1}$ of integrated luminosity \cite{ATLAS:2018wmg}. Furthermore, both collaborations have searched for the signal that would emerge from the processes $pp \to W_R \to Z W_L$ and $pp \to W_R \to H W_L$ in all possible final states (leptonic, semi-leptonic and hadronic), and derived that $M_{W_R} \ge 3.9$~TeV~\cite{CMS:2018ljc, ATLAS:2020fry}.

On different grounds, mass limits on $W_R$ bosons can also be obtained indirectly from low-energy constraints, especially from box diagrams contributing to kaon mixing. In the case where the CKM matrices in the right-handed and left-handed quark sectors are the same, we get $M_{W_R}>2.9$ TeV \cite{Zhang:2007fn}. Parity violation effects to be observed in polarised muon decays additionally impose that $M_{W_R}>600$~GeV~\cite{TWIST:2011egd}, and combined limits on $M_{W_R}$ and $M_{N_R}$ can be additionally derived~\cite{Prezeau:2003xn}. 

%%%%%%%%%%%%%%%%%%%%%%%%
\subsection{Production of $W_R$ bosons at the LHC}
\label{sec:xsections}
%%%%%%%%%%%%%%%%%%%%%%%%%
\begin{figure}
  \centering
  \includegraphics[width=0.4\linewidth]{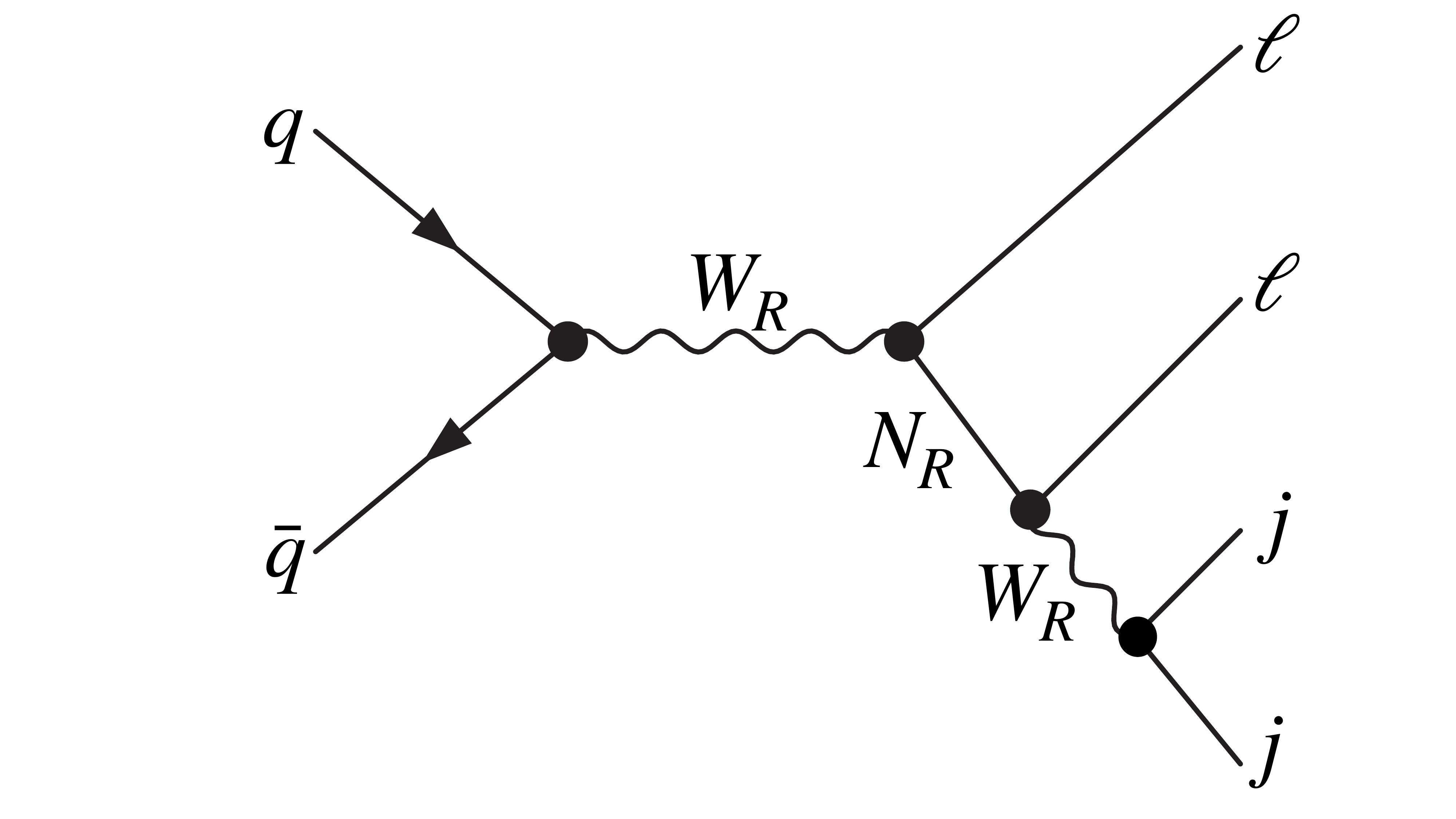}\hfill
  \includegraphics[width=0.4\linewidth]{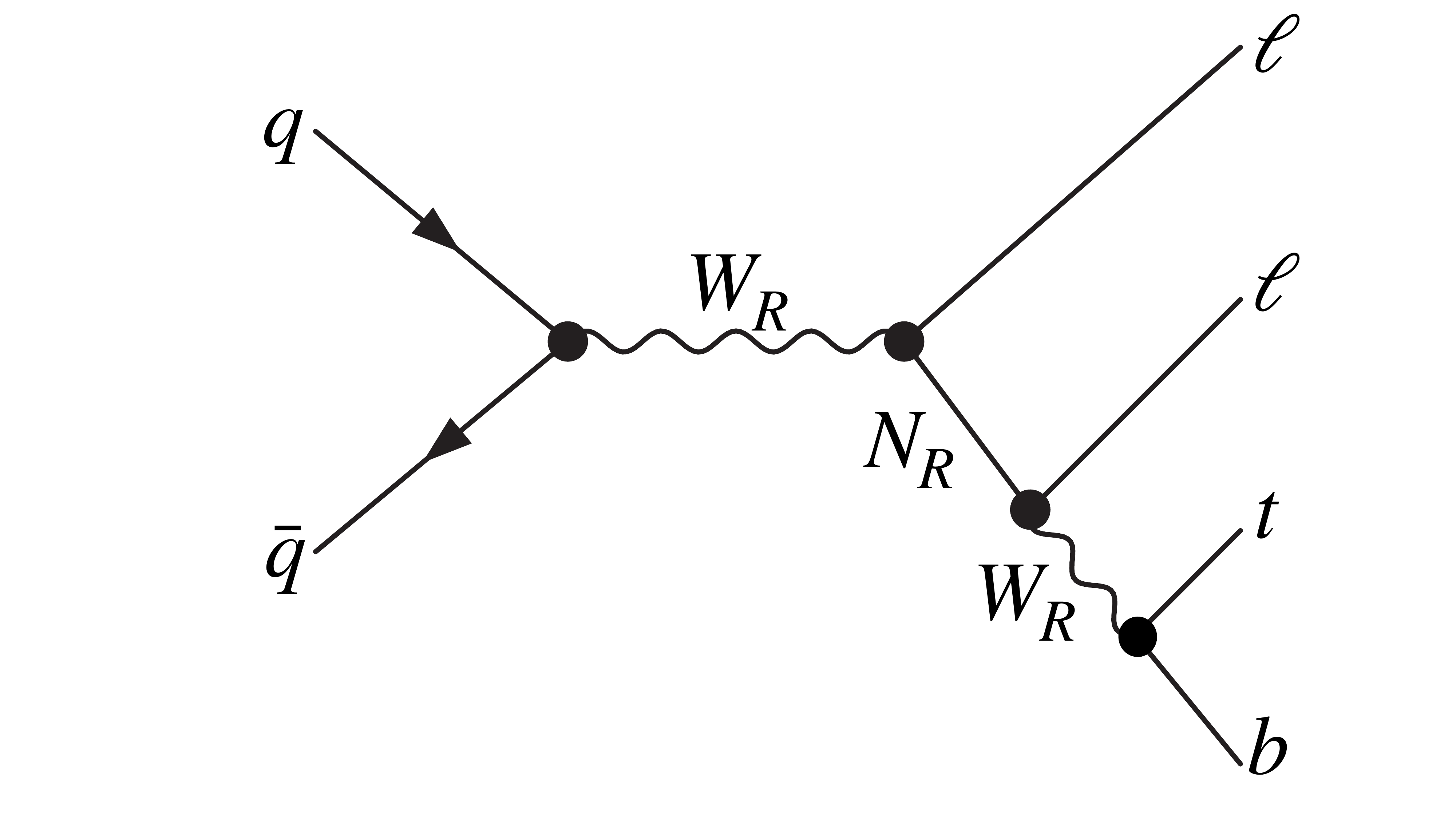}
  \caption{Representative parton-level Feynman diagrams illustrating the production of two leptons and either a pair of jets (left) or a top-bottom system (right) from the decay of a heavy right-handed neutrino $N_R$. The latter originates from the decay of a possibly off-shell $SU(2)_R$ charged gauge boson $W_R$ emerging from DY production.} \label{fig:FD}
\end{figure}

We consider the DY-like production of a (possibly off-shell) $W_R$ boson in $pp$ collisions, followed by its decay via the Keung-Senjanovic mechanism into two charged leptons and either two jets or a top-bottom system, 
\begin{equation}\label{eq:processes}
  p p \to W_R \to \ell N_R \to \ell \ell j j
  \qquad\text{or}\qquad
  p p \to W_R \to \ell N_R \to \ell \ell t b.
\end{equation}
In our notation, $j$ represents a light jet resulting from the fragmentation of quarks or antiquarks from the first or second generation, and $\ell$ stands for either an electron or a muon. Moreover, among all possible flavour and electric-charge assignments for the two leptons, we restrict our analysis to the case of a Same-Sign (SS) and Opposite-Sign (OS) lepton pair of the Same Flavour (SF). Representative parton-level Feynman diagrams illustrating the processes~\eqref{eq:processes} are shown in Figure~\ref{fig:FD}. In the following, we consider an mLRSM scenario in which the right-handed neutrino is lighter than the $W_R$ gauge boson ($M_{N_R} \leq M_{W_R}$), and we set the two CKM matrices to the $3\times 3$ identity matrix.

\begin{figure}
    \centering
    \includegraphics[width=0.49\linewidth]{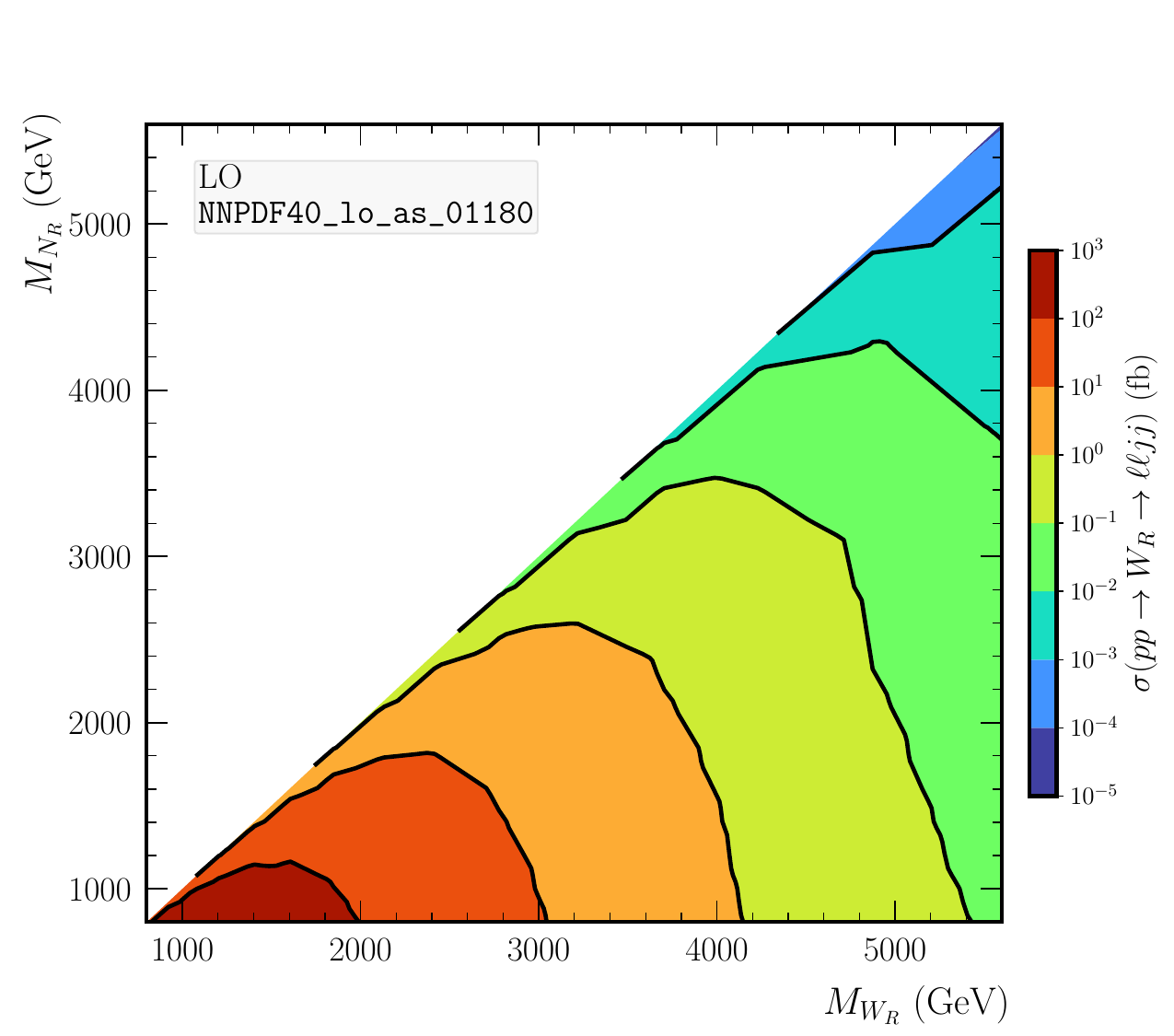}
    \hfill
    \includegraphics[width=0.49\linewidth]{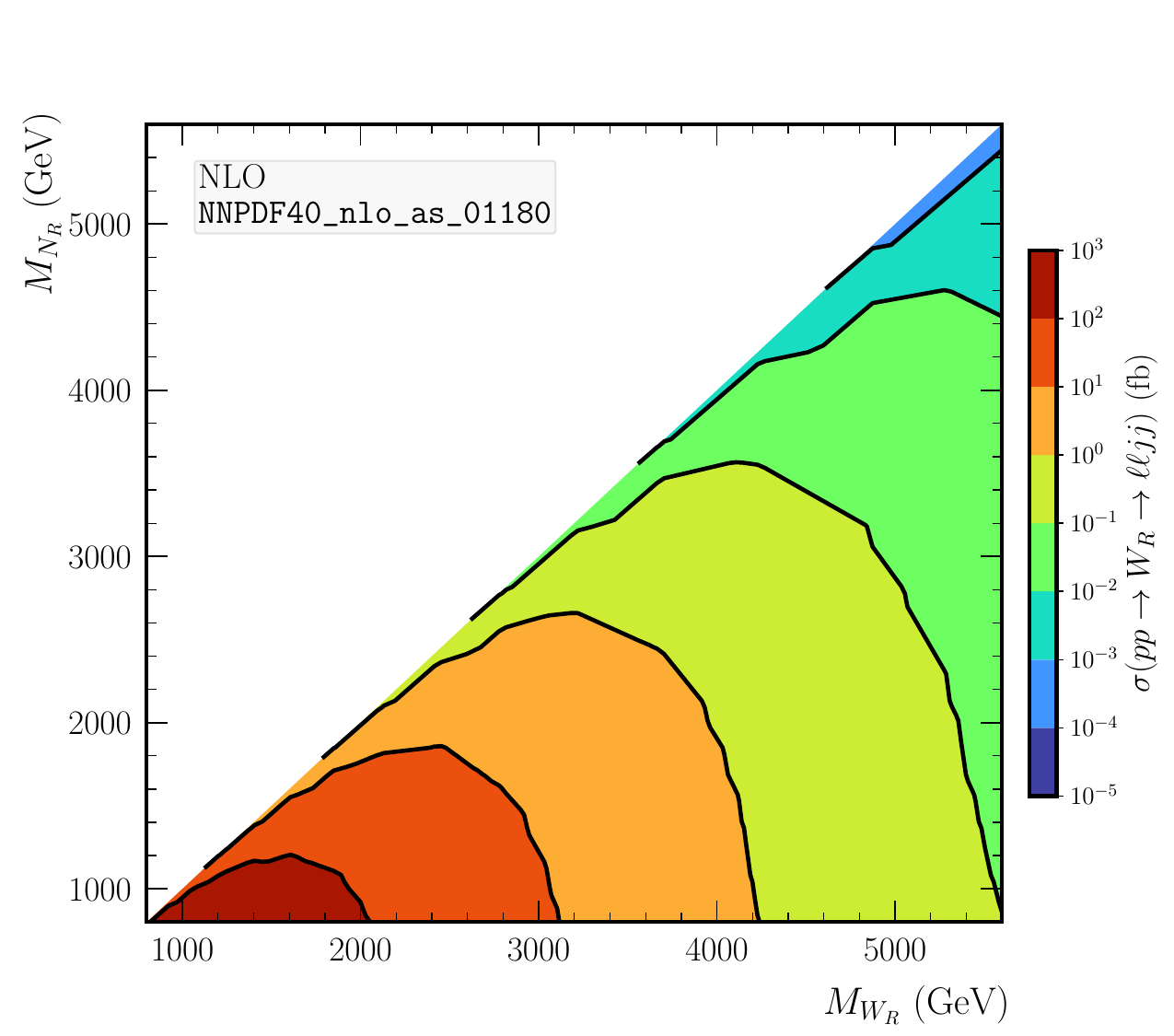}
    \vfill
    \includegraphics[width=0.49\linewidth]{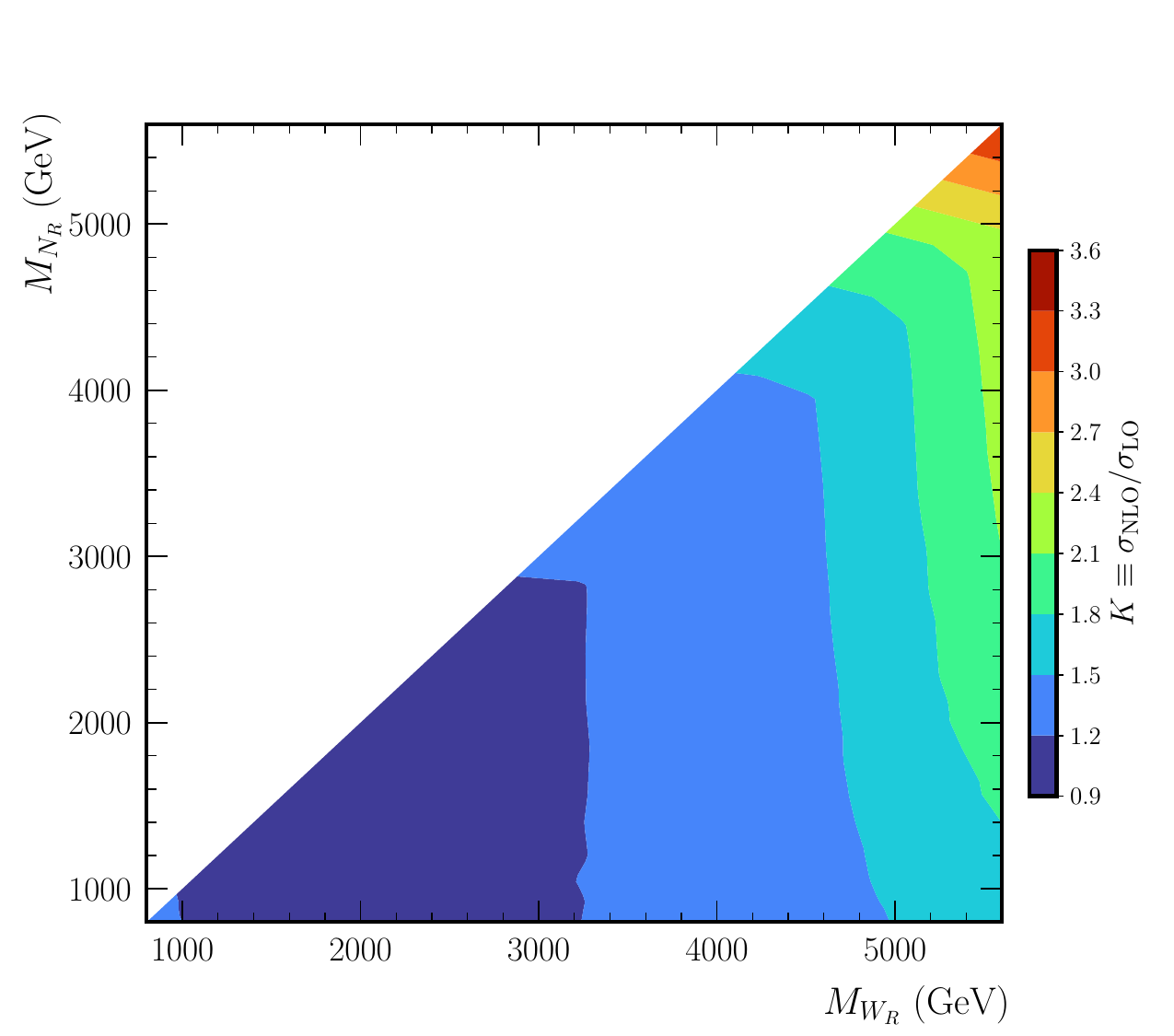}
    \caption{Production cross sections related to the process $pp\to W_R\to \ell N_R \to \ell\ell jj$ at LO (top row, left) and NLO (top row, right), for a centre-of-mass energy $\sqrt{s}=13$~TeV. We also display the $K$-factor defined as in~\eqref{eq:K} (bottom row).}
    \label{fig:XS:WR}
\end{figure}

We begin with a calculation of the cross section for the first of the processes~\eqref{eq:processes}, $pp\to\ell N_R\to \ell\ell jj$, both at Leading Order (LO) and Next-to-Leading Order (NLO) in QCD. We rely on the narrow-width approximation so that we can factorise the full process into a `production part' ($p p \to \ell N_R$) and a `decay part' ($N_R \to \ell jj$). The generation of the corresponding `production' LO and NLO matrix elements is achieved by means of \textsc{MadGraph5\_aMC@NLO} version 3.4.2 \cite{Alwall:2014hca} and a UFO~\cite{Degrande:2011ua,Darme:2023jdn} implementation of the relevant sectors of the mLRSM. The latter is obtained by means of \textsc{FeynRules}~\cite{Alloul:2013bka, Christensen:2009jx} and \textsc{NLOCT}~\cite{Degrande:2014vpa}, and an in-house effective implementation of the Lagrangian~\eqref{eq:lag} that considers only the $W_R$-boson interactions with fermions and the SM sector (on the same grounds as in the parametrisation proposed in Ref.~\cite{Fuks:2017vtl}). The `decay' matrix elements are instead always evaluated at LO thanks to \textsc{MadSpin}~\cite{Artoisenet:2012st}, which allows us to keep track of all spin correlations inherent to the fermionic nature of the right-handed neutrino $N_R$ and its decay products. Hadronic cross sections are next obtained by convoluting the resulting `full' matrix elements with the LO (\texttt{NNPDF40\_lo\_as\_01180}) and NLO (\texttt{NNPDF40\_nlo\_as\_01180}) sets of NNPDF~4.0 Parton Distribution Functions (PDFs)~\cite{NNPDF:2021njg} for LO and NLO calculations, respectively, and this relies on a central scale choice in which both the factorisation and renormalisation scales are fixed to the $W_R$-boson mass, $\mu_R = \mu_F = M_{W_R}$. To visualise the size of the NLO corrections, we define a global $K$-factor by the ratio of the LO to NLO rates $\sigma_{\rm LO} \equiv \sigma(pp \to \ell\ell jj)_{\rm LO}$ and $\sigma_{\rm NLO} \equiv \sigma(p p \to \ell \ell jj)_{\rm NLO}$,
\begin{equation}
  K \equiv \frac{\sigma_{\rm NLO}}{\sigma_{\rm LO}}.
\label{eq:K} \end{equation}

Production cross sections for the `full' process $pp\to\ell N_R\to \ell\ell jj$ are displayed in Figure~\ref{fig:XS:WR} at LO (top row, left) and NLO (top row, right), together with the associated $K$-factors as defined in~\eqref{eq:K} (bottom row). They are found to vary between 100--1000~fb for a relatively light $W_R$ boson with a mass between 1 and 2~TeV, to around $10^{-3}$ fb for a very heavy $W_R$ boson with a mass around 5~TeV, \textit{i.e.}\ in the vicinity of the exclusion limits relevant for a sequential extra charged gauge boson (the precise reinterpretation of these bounds in the mLRSM being addressed in section~\ref{sec:cmsbounds}). Furthermore, for a fixed value of the $W_R$-boson mass, the cross section value depends on the mass splitting between the $W_R$ and $N_R$ states. It hence decreases with decreasing values of $\Delta \equiv M_{W_R} - M_{N_R}$, as expected from the reduced phase space available for the decay when the two masses are similar. We also recall that cross sections for the electron and muon channels are identical, as predicted by the universality of the $SU(2)_R$ gauge coupling in the Lagrangian of Eq.~\eqref{eq:lag}. Finally, $K$-factors are modest for most of the mass regime considered (\textit{i.e.}\ for $M_{W_R} < 5$~TeV), and they lie between 1.1 and 1.5. The lowest values correspond to lighter $W_R$ boson mass configurations while the larger ones correspond to the heaviest setups considered. Such a result is typical of the DY-like production of extra gauge bosons with masses in the TeV range, recalling that the behaviour of the $K$-factor is mostly independent of how the central scale is chosen~\cite{Sullivan:2002jt, Fuks:2017vtl}. The $K$-factor however further increases to up to 3.3 for heavier $W_R$ bosons with masses above 5~TeV, that also correspond to a regime in which PDF uncertainties are much larger and yield larger differences between LO predictions (involving a LO set of parton densities often associated with a poorer fit to data) and NLO predictions (involving an NLO set of parton densities).

%%%%%%%%%%%%%%%%%%%%%%%%%%%%%%%
\subsection{Searches for $W_R$ bosons in events with two leptons and two jets}\label{sec:cmsbounds}
%%%%%%%%%%%%%%%%%%%%%%%%%%%%%%%
In the previous sections, we mentioned the potential bounds that could be extracted from the first of the processes shown in Eq.~\eqref{eq:processes} and the associated searches at the LHC. In this section, we study the consequences of such searches on the model. To this aim, we reinterpret the results of the CMS-EXO-20-002 search targeting final states comprising two leptons (\textit{i.e.}\ electrons or muons) and two (light) jets, using data collected between 2016 and 2018 and corresponding to an integrated luminosity of $138~{\rm fb}^{-1}$~\cite{CMS:2021dzb}. However, instead of extracting constraints directly from the published experimental results, we undertake our own interpretation study. The reason is that in the initial analysis, the CMS collaboration simulated the signal at LO, used global $K$-factors (as computed in section~\ref{sec:xsections}), and followed the prescription of Refs.~\cite{Mattelaer:2016ynf,Mitra:2016kov}. Instead, we perform a more precise signal simulation at NLO in QCD, including real and virtual contributions to ${\cal O}(\alpha_{\rm s})$ at the fully differential level, and we additionally rely on more recent PDF sets. Our reinterpretation study however ignores any potential correlation between the different signal regions of the analysis, as relevant information has not been released by the CMS collaboration. In light of our findings and the experimental results published, we however do not expect that this issue would yield any significant change in our conclusions. 

\begin{table}
\setlength\tabcolsep{10pt}\renewcommand{\arraystretch}{1.2}
\begin{center}
\begin{tabular}{l cc cc}
& \multicolumn{2}{ c }{Electron channel} & \multicolumn{2}{c}{Muon channel} \\
$]M_{\ell\ell jj}^{\rm min}, M_{\ell\ell jj}^{\rm max}]$ & Data & Background &  Data & Background \\ 
\toprule
$]800,1000]$ GeV& $1106.0$ & $1103.5\pm 26.607$  & $1639.0$ & $1670.7 \pm 39.774$  \\
$]1000,1200]$ GeV& $646.0$ & $631.51\pm 16.968$ & $946.0$ & $925.99 \pm 23.917$ \\ 
$]1200,1400]$ GeV& $332.0$ & $323.23\pm10.736$ & $518.0$ & $500.33 \pm 14.869$ \\ 
$]1400,1600]$ GeV& $170.0$ & $169.69\pm6.8418$ & $268.0$ & $263.88 \pm 9.3498$ \\
$]1600,2000]$ GeV& $143.0$ & $157.55\pm9.505$ & $216.0$ & $215.18 \pm 8.2146$  \\
$]2000,2400]$ GeV& $62.0$ & $52.327\pm3.9676$ & $80.0$ & $73.482 \pm 4.4654$ \\ 
$]2400,2800]$ GeV& $25.0$ & $19.567\pm1.5493$  & $30.0$ & $25.943 \pm 2.3125$ \\
$]2800,3200]$ GeV& $10.0$ & $8.9907\pm 1.209$  & $13.0$ & $9.7557 \pm 1.1603$ \\
$]3200,8000]$ GeV & $13.0$ & $6.2463\pm0.77892$  & $11.0$ & $7.8119 \pm 0.84286$ \\ 
\end{tabular}
\end{center}
\caption{Definition of the 18 signal regions of the CMS-EXO-20-002 search for $W_R$-boson production and decay in the $\ell\ell jj$ channel~\cite{CMS:2021dzb}. After a common pre-selection (see the description in the text), the different regions are defined according to the lepton flavour (electron or muon) and the value of the reconstructed $W_R$-boson mass $M_{\ell\ell jj}$. For each region, we show the number of observed events and the associated SM expectation.}
\label{tab:CMS:SRs}
\end{table}

The CMS-EXO-20-002 search is dedicated to the analysis of events featuring exactly two isolated leptons (electrons or muons, regardless of their electric charge), and at least two jets reconstructed by means of the anti-$k_T$ algorithm~\cite{Cacciari:2008gp, Cacciari:2011ma} with a radius parameter fixed to $R=0.4$. For events exhibiting more than two jets, the two leading jets in terms of transverse momentum are considered to originate from the decay of a $W_R$ boson, together with the two leptons. The leading and sub-leading charged leptons are required to have transverse momenta $p_T > 60$ and 53~GeV respectively, and to be within the detector acceptance (\textit{i.e.}\ with a pseudo-rapidity $|\eta| < 2.4$, excluding electrons and positrons with a pseudo-rapidity lying in the interval $1.44 < |\eta| < 1.57$ that corresponds to the transition region between the barrel and endcap of the CMS electromagnetic calorimeter). Furthermore, selected jets must have a transverse momentum $p_T > 40$~GeV and pseudo-rapidity $|\eta| < 2.4$, the invariant mass of the reconstructed $W_R$-boson candidate is enforced to verify $M_{\ell\ell jj} > 800$~GeV, and the invariant mass of the lepton pair is constrained to fulfill $M_{\ell\ell} > 400$~{\rm GeV}. After this pre-selection, the analysis defines 18 signal regions, 9 in the $eejj$ channel and 9 in the $\mu\mu jj$ channel, the various regions corresponding in different bins in $M_{\ell\ell jj}$. We report the exact definition of these bins, together with their background expectations and the associated observations, in Table~\ref{tab:CMS:SRs}.

In order to determine the constraints that could be imposed on the mLRSM from LHC searches for $W_R$ signatures in the $\ell\ell j j$ channel, we have implemented the CMS-EXO-20-002 search in the \textsc{MadAnalysis}~5 framework~\cite{Conte:2012fm, Conte:2014zja, Conte:2018vmg}. Our implementation relies on the SFS module for the simulation of the detector effects via a parametrisation through transfer functions~\cite{Araz:2020lnp, Araz:2021akd}, and it has been thoroughly validated by ensuring an excellent agreement with predictions provided by the CMS collaboration on \textsc{HepData}~\cite{hepdata.114866}. Details on our implementation and its validation can be found in Appendix~\ref{sec:MA5:CMS}, as well as on the Public Analysis Database (PAD)~\cite{Dumont:2014tja} and the dataverse~\cite{DVN/UMGIDL_2023} of \textsc{MadAnalysis}~5. In addition, we also provide in this appendix information on the implementation and validation of the superseded CMS-EXO-17-011 search dedicated to the same signal, but in which only a partial LHC Run~2 dataset is analysed. This older search had been initially used for the present work, until the more recent CMS-EXO-20-002 results appeared.

\begin{figure}
\centering
\includegraphics[width=0.49\linewidth]{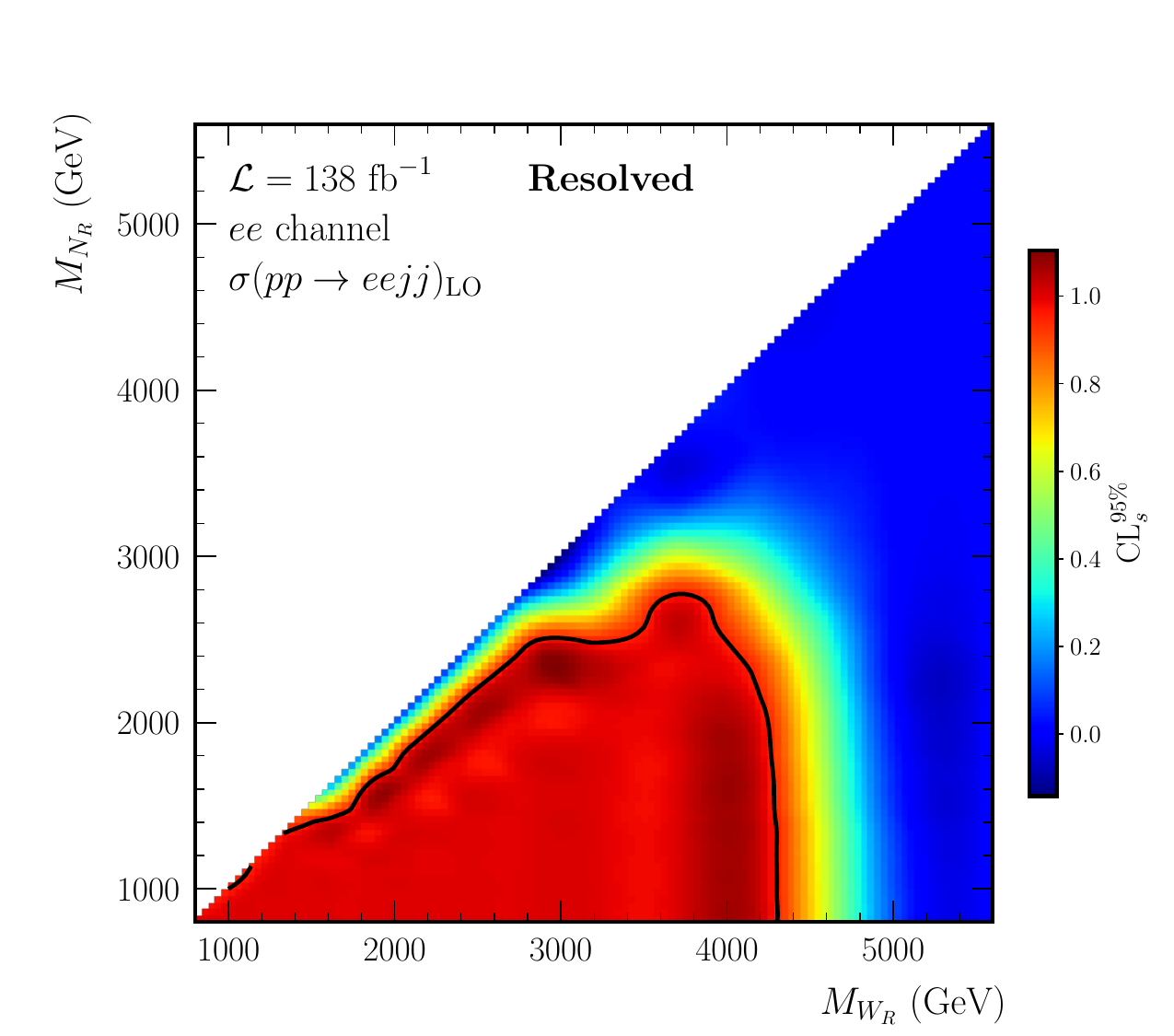}
\hfill
\includegraphics[width=0.49\linewidth]{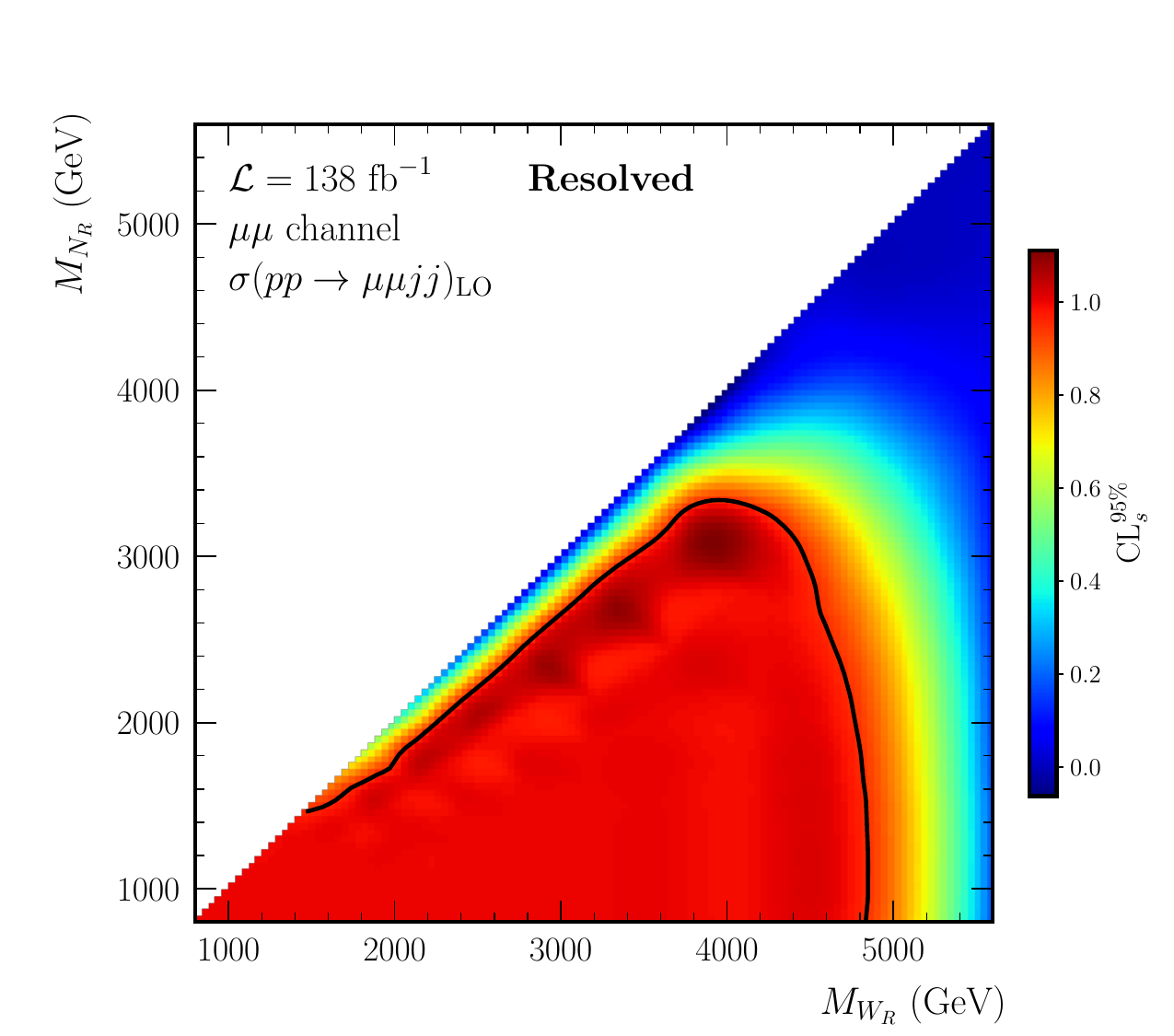}
\vfill
\includegraphics[width=0.49\linewidth]{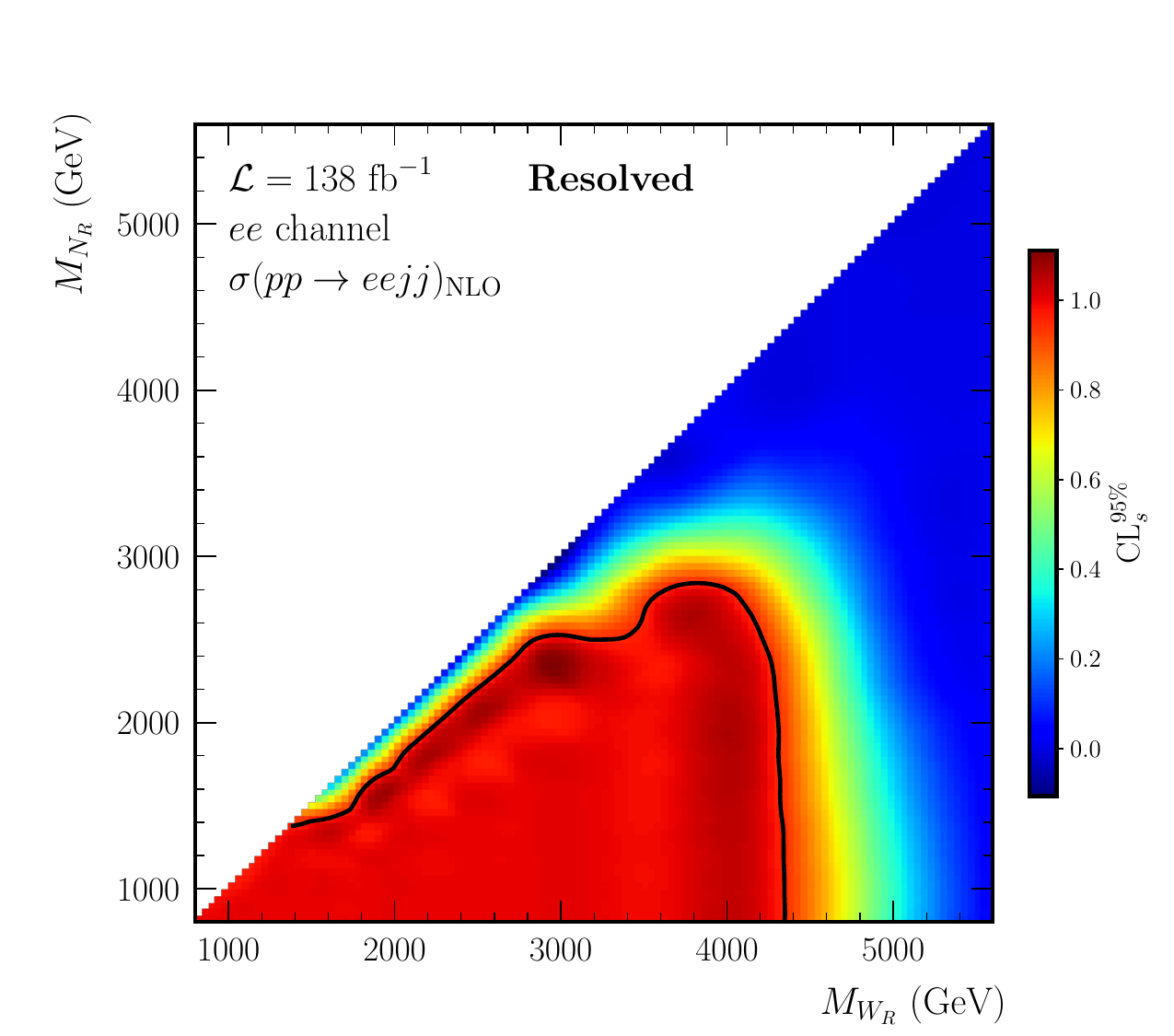}
\hfill
\includegraphics[width=0.49\linewidth]{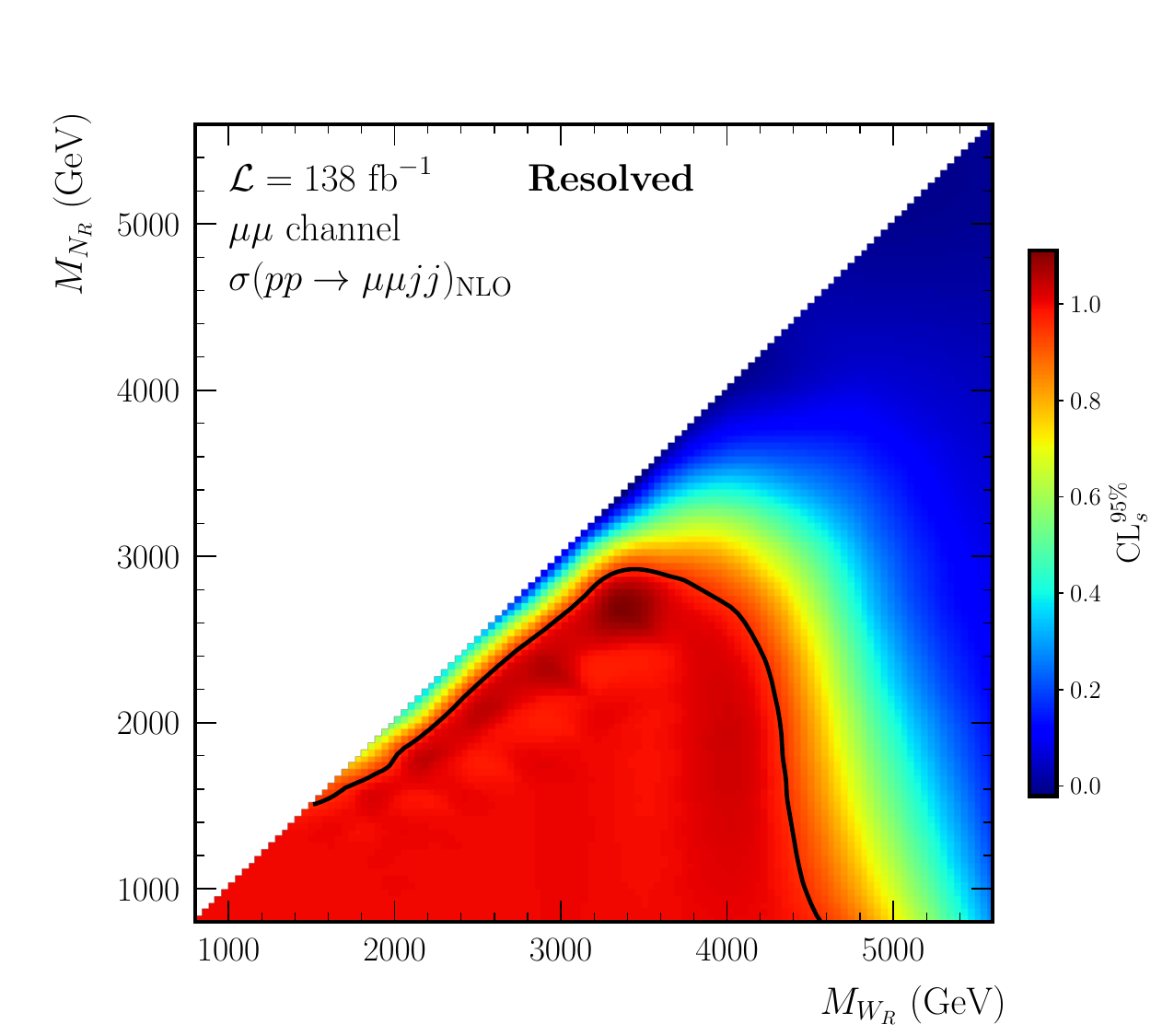}
\caption{Exclusion contours extracted from the reinterpretation of the results of the CMS-EXO-20-002 analysis, projected on the plane ($M_{W_R}$, $M_{N_R}$) in the electron (left) and muon (right) channel, when using matrix elements calculated at LO (top row) and NLO (bottom row). The black solid line corresponds to a CL$_s$ value of $1-{\rm CL}_s=0.95$, all mass points localised on its lower left being thus excluded.}
\label{fig:CLs:LHC}
\end{figure}

The resulting constraints on the model are derived from a scan of the two-dimensional plane defined by the masses $M_{W_R}$ and $M_{N_R}$, which we vary in the range [800, 5300]~GeV. We analysed 120 signal configurations with $M_{N_R} \leq M_{W_R}$, generating 500,000 hard-scattering events for each. Subsequently, we utilised \textsc{Pythia} version 8.309~\cite{Bierlich:2022pfr} for simulating parton showering and hadronisation. The resulting hadron-level events were then analysed with \textsc{MadAnalysis}~5, employing the CL$_s$ method~\cite{Read:2002hq} to assess the signal's viability with respect to data. We report the obtained CL$_s$ values in Figure~\ref{fig:CLs:LHC}, both for the electron (left) and muon (right) channels and using LO (top row) and NLO (bottom row) simulations. We obtain in general results in the same ballpark as those presented officially in the CMS analysis~\cite{CMS:2021dzb}, demonstrating hence by different means the validity of our implementation of the CMS analysis in our tool chain.

In the electron channel, we exclude $W_R$ bosons with masses smaller than 4.3~TeV, given that the mass splitting $\Delta$ between the $W_R$ boson and the right-handed neutrino $N_R$ is significant. In the most extreme situation featuring a $W_R$ boson of 4--5~TeV, $N_R$ masses ranging up to 3--3.2~TeV can be reached. The CMS official bounds are, in comparison, slightly stronger by about 1-2$\sigma$. This difference is mitigated when comparing our NLO exclusions to the CMS official ones, which are based on LO simulations including an NLO $K$-factor. In this case, the remaining difference is less than 10\% when considering the mass limit values, as expected due to the entirely different nature of the detector simulations employed in this work (the SFS framework) and in the CMS analysis (the CMS simulation software based on \textsc{Geant}~4).

\begin{figure}
    \centering
    \includegraphics[width=0.49\linewidth]{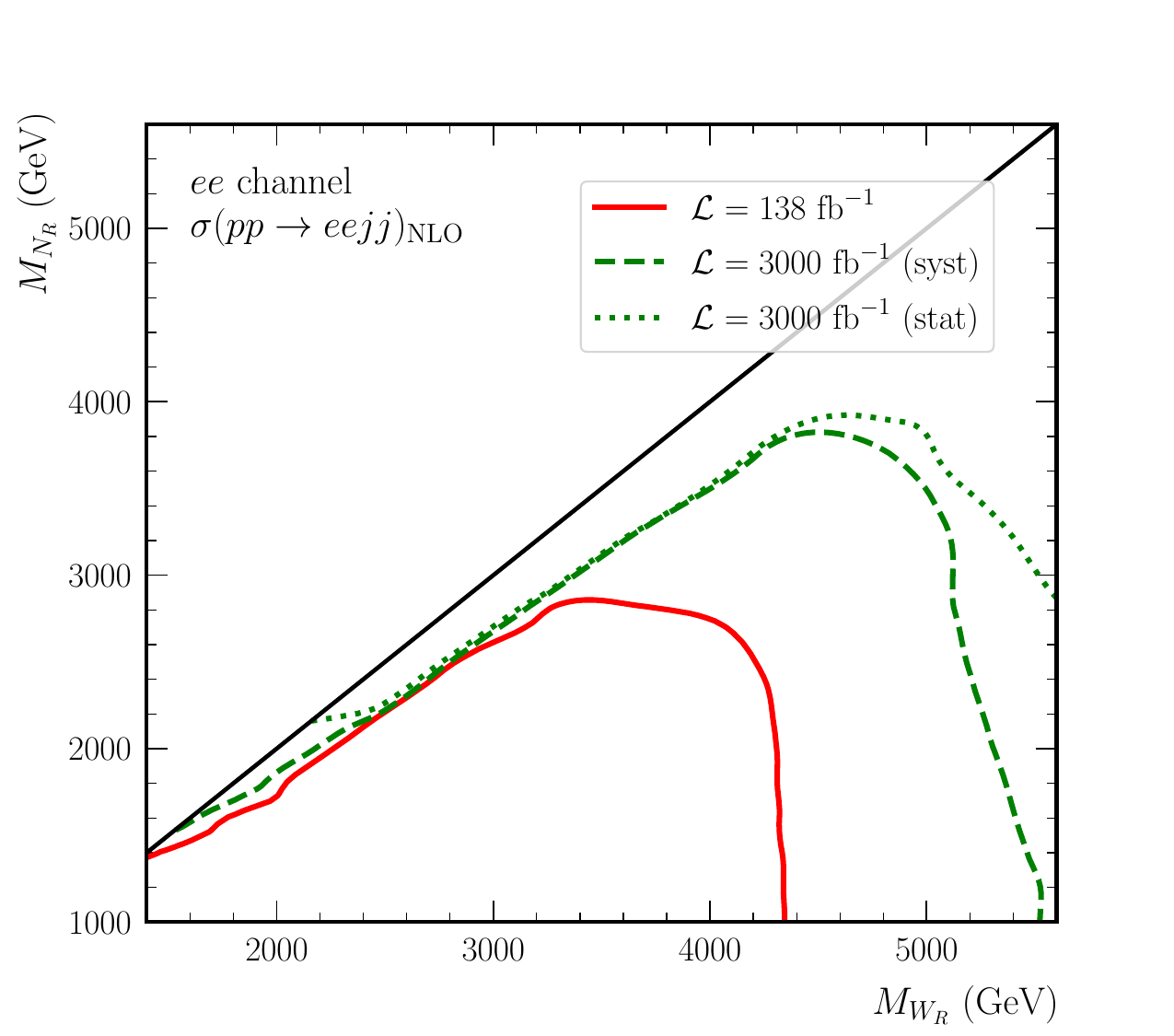}
    \includegraphics[width=0.49\linewidth]{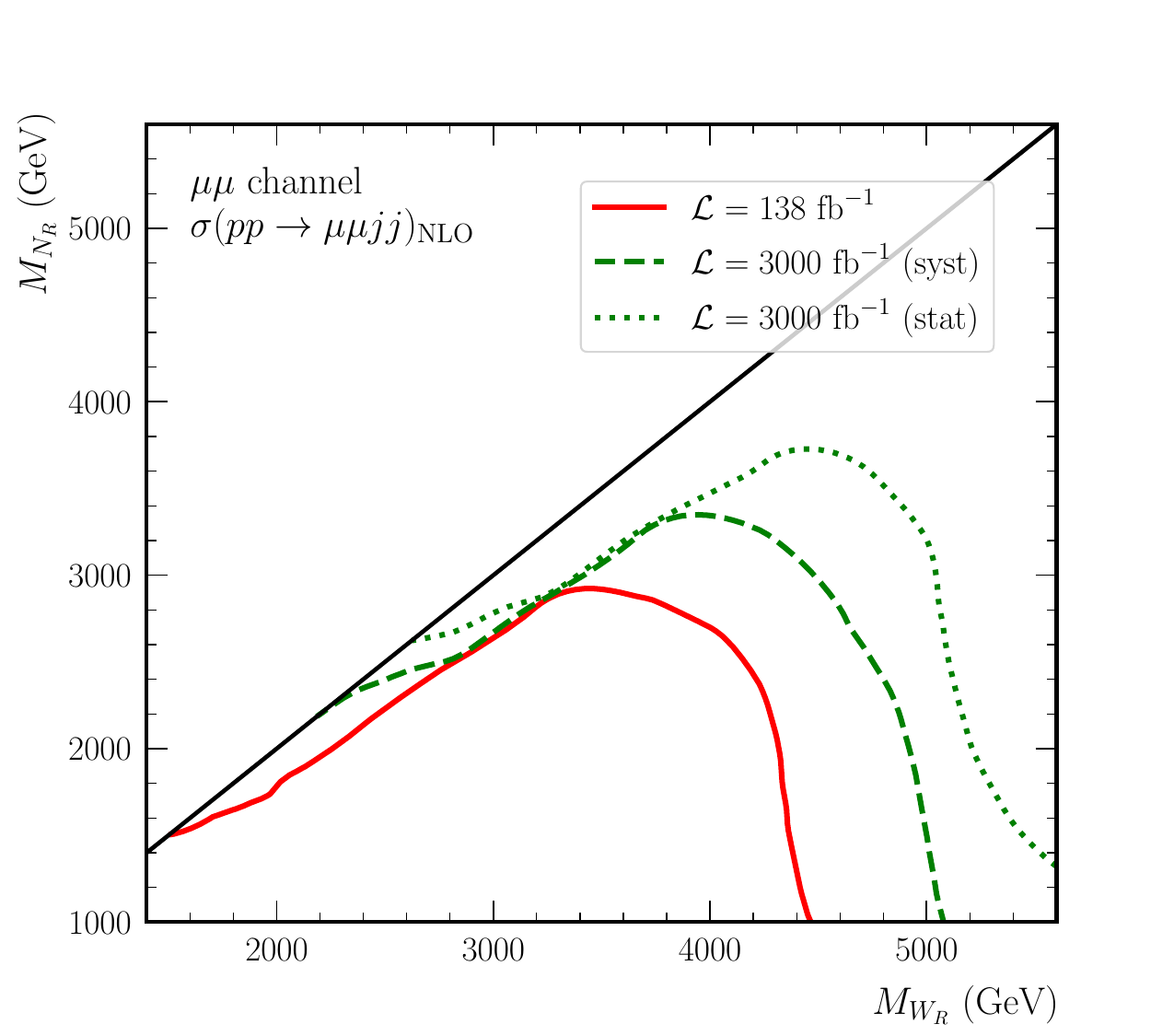}
    \caption{Exclusion contours in the plane $(M_{W_R}, M_{N_R})$ as obtained from the recasting of the CMS-EXO-20-002 search in the $eejj$ (left) and $\mu\mu jj$ (right) channel, from NLO signal simulations. We compare current exclusions (solid red) to expectations at the high-luminosity LHC with 3000~fb$^{-1}$,  assuming that the background uncertainties are dominated by the statistical (dotted green) or systematic (dashed green) components.}
    \label{fig:CLs:expected}
\end{figure}

Results in the muonic channel are more intriguing. Similar to the electronic case, our LO limits are in the same ballpark as the CMS official ones, with the latter being slightly stronger. However, with NLO simulations we observe a decrease in sensitivity predicted by \textsc{MadAnalysis}~5. It is important to note that the limits found lie close to the regime where the $K$-factors displayed in Figure~\ref{fig:XS:WR} are larger than 2, in contradiction with naive expectations of a DY-like process. This behaviour can be traced back to the impact of the differences between the poor LO and the better NLO NNPDF~4.0 parton density fits, which therefore significantly modify not only total rates but also event kinematics. As NLO PDF fits are better than LO fits, we rely on NLO simulations only from now on.

We close this section by estimating the change in the exclusion contours when the LHC luminosity is scaled to ${\cal L} = 3000~{\rm fb}^{-1}$, which corresponds to the high-luminosity LHC phase. We assume the same search strategy as in~\cite{CMS:2021dzb}, and naively re-scale the exclusion contours following the methods presented in Ref.~\cite{Araz:2019otb}. We consider that the expected number of background events for each signal region, shown in Table~\ref{tab:CMS:SRs}, scales proportionally to the luminosity, and that the scaling of the uncertainties on the background is achieved in two different complementary ways depending on whether background errors are dominated by statistical uncertainties or by systematic ones. We consider the two cases, and derive expected limits by assuming observations in agreement with the SM background. The results are displayed in Figure~\ref{fig:CLs:expected}. They show that $W_R$-boson masses between 5 and 6 TeV can be reached, together with right-handed neutrino masses ranging up to 4 TeV.
%%%%%%%%%%%%%%%%%%%%%%%%%%%%%%%%%%%%%%%%%%%%%%%%%%%%%%%%%%%%%%%%%%%%%%%%%%%
\section{Sensitivity of $\ell\ell tb$ probes to left-right models}
\label{sec:tbll}
%%%%%%%%%%%%%%%%%%%%%%%%%%%%%%%%%%%%%%%%%%%%%%%%%%%%%%%%%%%%%%%%%%%%%%%%%%%
In section~\ref{sec:theo}, we focused on existing LHC searches for mLRSM signals in which a $W_R$ boson decays into a lepton pair and a di-jet system. However, for heavier spectra, decays into a lepton pair and a top-bottom system could as well contribute. We therefore focus in this section on the second process of Eq.~\eqref{eq:processes}, illustrated by the second diagram in Figure~\ref{fig:FD}. In order to assess the potential of the LHC to such a signal, we consider a set of benchmark scenarios that satisfy all current constraints, and then design a dedicated phenomenological analysis that could potentially seed future experimental searches in proton-proton collisions at a centre-of-mass energy of $\sqrt{s}=13.6$~TeV.

%%%%%%%%%%%%%%%%%%%%%%%%%%%%%%%%%%%%%
\subsection{Benchmark scenarios and backgrounds}\label{sec:bench_bgd}
%%%%%%%%%%%%%%%%%%%%%%%%%%%%%%%%%%%%%

\begin{table}
\setlength\tabcolsep{18pt}
\begin{center}
\begin{tabular}{l ccc}
BP & BP$e$1 & BP$e$2 & BP$e$3 \\
\toprule
$M_{W_R}$   [GeV] & 4800 & 4800 & 4800 \\
$M_{N_R}$   [GeV] & 960  & 2400 & 4400 \\
$\sigma(pp \to \ell\ell tb)_{\rm LO}$ [fb] & $1.22 \times 10^{-1}$ & $7.77 \times 10^{-2}$ & $4.27 \times 10^{-3}$ \\
$\sigma(pp \to \ell\ell tb)_{\rm NLO}$ [fb] & $1.73 \times 10^{-1}$ & $1.13 \times 10^{-1}$ & $6.43 \times 10^{-3}$ \\
$\Gamma_{W_R}~[{\rm GeV}]$ & $134$ & $130$ & $122$  \\
$\Gamma_{N_R}~[{\rm GeV}]$ & $2.47 \times 10^{-5}$ & $2.96 \times 10^{-3}$ & $1.21 \times 10^{-1}$ \\[.4cm]
BP & BP$e$4 & BP$e$5 & BP$e$6 \\ 
\toprule
$M_{W_R}$   [GeV] & 5500 & 5500 & 5500 \\
$M_{N_R}$   [GeV] & 1100 & 2750 & 5100 \\
$\sigma(pp \to \ell\ell tb)_{\rm LO}$ [fb] & $3.39 \times 10^{-2}$ & $1.73 \times 10^{-2}$ & $6.87 \times 10^{-4}$ \\
$\sigma(pp \to \ell\ell tb)_{\rm NLO}$ [fb] & $5.50 \times 10^{-2}$ & $3.13 \times 10^{-2}$ & $1.34 \times 10^{-3}$ \\
$\Gamma_{W_R}~[{\rm GeV}]$ & $153$ & $149$ & $139$ \\
$\Gamma_{N_R}~[{\rm GeV}]$ & $2.88 \times 10^{-5}$ & $3.41 \times 10^{-3}$  & $1.52 \times 10^{-1}$\\[.4cm]
BP & BP$\mu$1 & BP$\mu$2 & BP$\mu$3 \\
\toprule
$M_{W_R}$   [GeV] & 5100 & 5100 & 5100 \\
$M_{N_R}$   [GeV] & 1020 & 2550 & 4700 \\
$\sigma(pp \to \ell\ell tb)$ [fb] & $6.99 \times 10^{-2}$ & $4.09 \times 10^{-2}$ & $1.54 \times 10^{-3}$ \\
$\sigma(pp \to \ell\ell tb)_{\rm NLO}$ [fb] & $1.04 \times 10^{-1}$ & $6.45 \times 10^{-2}$ & $3.25 \times 10^{-3}$ \\
$\Gamma_{W_R}~[{\rm GeV}]$ & $142$ & $138$ & $129$ \\
$\Gamma_{N_R}~[{\rm GeV}]$ & $2.65 \times 10^{-5}$ & $3.15 \times 10^{-3}$ & $1.34 \times 10^{-1}$ \\[.4cm]
BP & BP$\mu$4 & BP$\mu$5 & BP$\mu$6 \\ 
\toprule
$M_{W_R}$   [GeV] & 5500 & 5500 & 5500 \\
$M_{N_R}$   [GeV] & 1100 & 2750 & 5100 \\
$\sigma(pp \to \ell\ell tb)_{\rm LO}$ [fb] & $3.39 \times 10^{-2}$ & $1.73 \times 10^{-2}$ & $6.87 \times 10^{-4}$ \\
$\sigma(pp \to \ell\ell tb)_{\rm NLO}$ [fb] & $5.50 \times 10^{-2}$ & $3.13 \times 10^{-2}$ & $1.34 \times 10^{-3}$ \\
$\Gamma_{W_R}~[{\rm GeV}]$ & $153$ & $149$ & $139$ \\
$\Gamma_{N_R}~[{\rm GeV}]$ & $2.88 \times 10^{-5}$ & $3.41 \times 10^{-3}$  & $1.52 \times 10^{-1}$\\
\end{tabular}
\end{center}
\caption{Definition of the 12 BPs used in our analysis, for $W_R$ decays into a $eetb$ system (BP$e$1 to BP$e$6) and $\mu\mu tb$ system (BP$\mu$1 to BP$\mu$6). These benchmarks satisfy current constraints originating from $W_R$ production and decay into an $\ell\ell jj$ final state, and they feature a variety of split and more compressed spectra. For each scenario, we additionally report LO and NLO signal cross sections at $\sqrt{s}=13.6$ TeV, as well as the total decay widths of the $W_R$ and $N_R$ states.}
\label{tab:BPs}
\end{table}

In order to build our analysis, we define Benchmark Points (BPs) that fulfill all constraints from searches for $W_R$ bosons in the $\ell\ell jj$ channel presented in the previous section. We consider two mass values for a $W_R$ boson decaying in the di-electron channel ($W_R\to eetb$), namely $M_{W_R}=4800$ and $5500$ GeV, and two mass values for the di-muon channel ($W_R\to \mu\mu tb$), namely $M_{W_R}=5100$ and $5500$ GeV. For each of these mass values, we then define three scenarios differing by the choice of the right-handed neutrino mass,
$$
M_{N_R} \in \bigg\{ \frac{M_{W_R}}{5},\ \frac{M_{W_R}}{2},\ M_{W_R} - 400~{\rm GeV}\bigg\}.
$$ 
Such a choice allows us to span most of the possible kinematic configurations in terms of the splitting $\Delta \equiv M_{W_R} - M_{N_R}$. The definitions of these different BPs are collected in Table~\ref{tab:BPs} for both the electron channel (upper part of the table) and the muon channel (lower part of the table), and we also indicate some of the properties of these scenarios relevant for the signal.

\begin{table}
\setlength\tabcolsep{10pt}\renewcommand{\arraystretch}{1.2}
\begin{center}
\begin{tabular}{l cccc}
Process & $\sigma_{\rm LO}$~[fb] & $\sigma_{\rm NLO}$~[fb] & $K \equiv \sigma_{\rm NLO}/\sigma_{\rm LO}$ & $N_{\rm events}$ \\
\toprule
$p p \to t ~\bar{t}~H$  & $345.5$ & $503.8$ & $1.45$ & $1.6 \times 10^6$ \\
$p p \to t ~\bar{t}~Z$ & $519.1$ & $838.9$ & $1.61$ & $1.6 \times 10^6$ \\
$p p \to t ~\bar{t}~W^\pm$ & $434.0$ & $670.2$ & $1.54$ & $1.6 \times 10^6$ \\
$p p \to t ~Z j + {\rm c.c.}$~\cite{Campbell:2013yla}  & $821.4$ & $903.5$  & $1.10$ & $2.2 \times 10^6$\\
$p p \to t~ W^- \bar{b} + {\rm c.c.}$~\cite{Demartin:2016axk} & $976.4$ & $1331.5$ & $1.36$  &  $4.0 \times 10^6$\\
\end{tabular}
\end{center}
\caption{Total cross sections at $\sqrt{s}=13.6$ TeV for all background processes relevant to our study, shown both at LO and NLO, along with the corresponding $K$-factor and the number of   generated events. Here, $\sigma_{tZj} \equiv \sigma(pp \to tZj) \times {\rm BR}(t\to bjj) \times {\rm BR}(Z\to \ell^+\ell^-)$ and $\sigma_{tWb} \equiv \sigma(pp \to tWb) \times {\rm BR}(t \to b \ell\nu) \times {\rm BR}(W \to \ell \nu)$. Total rate predictions have been obtained with \textsc{MadGraph5\_aMC@NLO}, unless indicated otherwise through a reference.}
\label{tab:bkg:xs}
\end{table}

In order to extract the mLRSM signal from the SM background, we focus on hadronic top quark decays so that the final state considered comprises exactly two charged leptons, and several jets whose hardness and multiplicity depend on the kinematic properties of the signal. For relatively small top quark transverse momenta (typically smaller than 500~GeV), events generally feature at least four small-radius jets, while for larger top quark transverse momenta we should have at least one large-$R$ jet and at least one small-$R$ jet. The lowest $N_R$ mass chosen for the scenarios explored in our study being fixed to $960$~GeV (BP$e$1), the produced top quarks have typically very high transverse momenta in most of the targeted cases. mLRSM spectra yielding softer top quarks are indeed already challenged by existing data and thus hard to justify phenomenologically. The backgrounds associated to our signal process thus usually emerge from the production of high-$p_T$ top quarks, with a much smaller contribution originating from the production of two massive gauge bosons in association with jets that we therefore do not consider. The list of all background contributions is given in Table~\ref{tab:bkg:xs}, together with the related LO and NLO cross sections.

\begin{figure}
  \centering
  \includegraphics[width=0.47\linewidth,trim={20 0 35 0},clip]{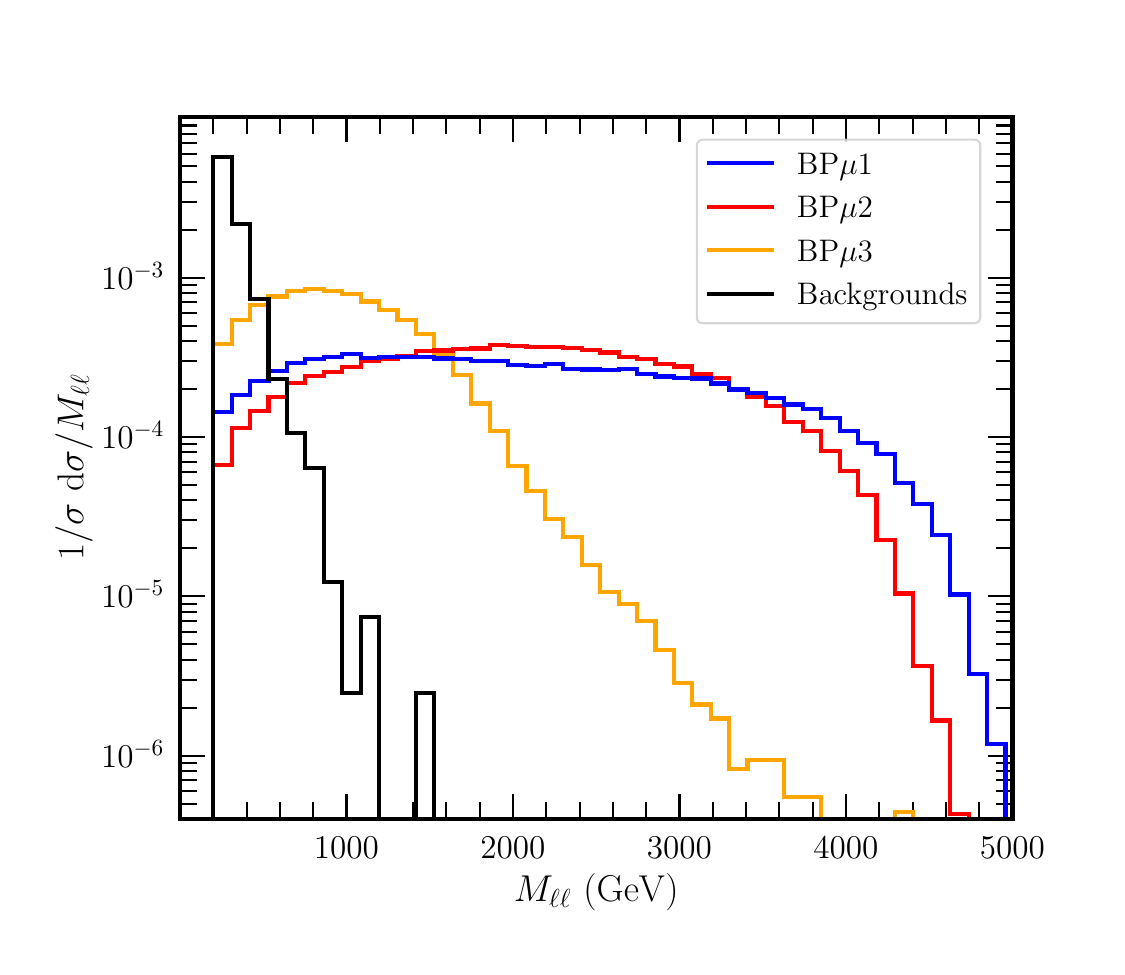}\hfill
  \includegraphics[width=0.47\linewidth,trim={20 0 35 0},clip]{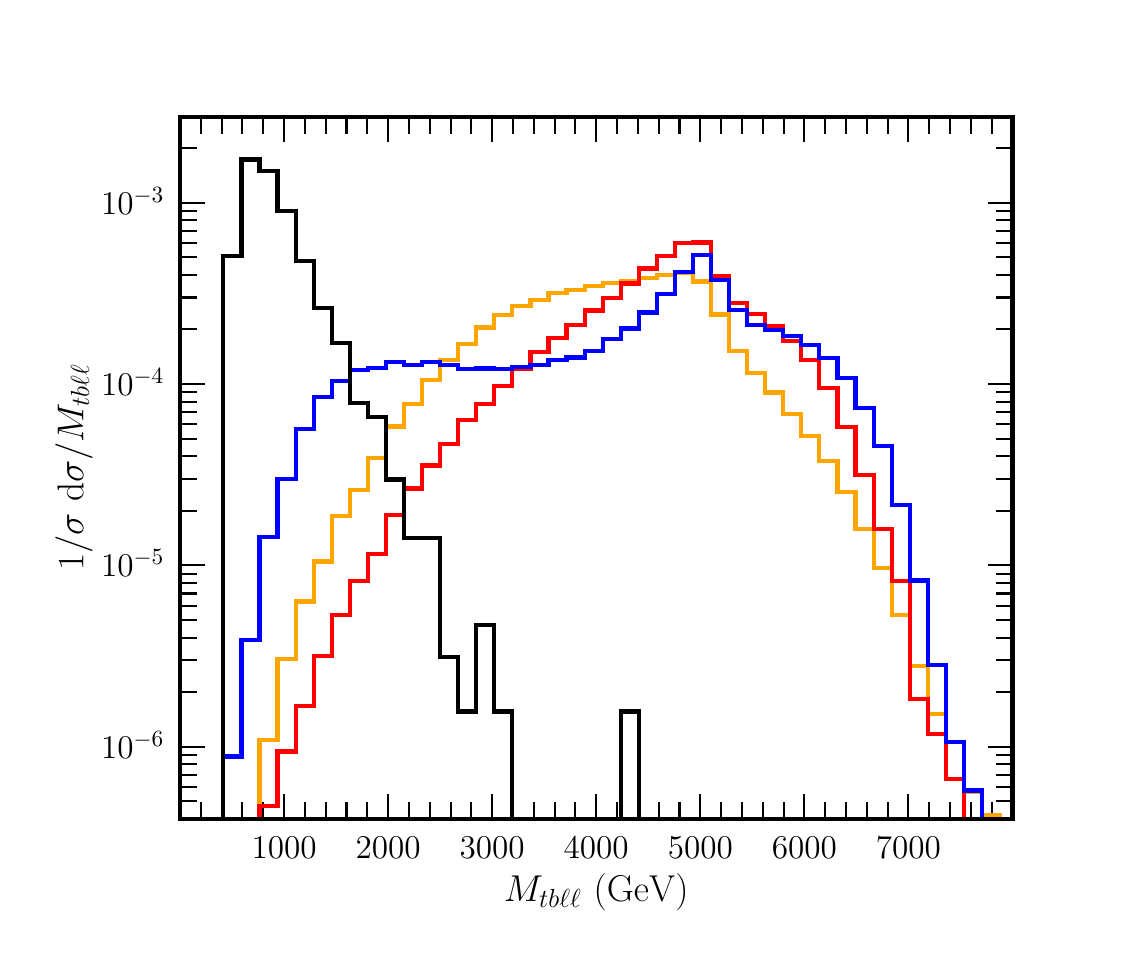}\\
  \includegraphics[width=0.45\linewidth,trim={7 0 60 0},clip]{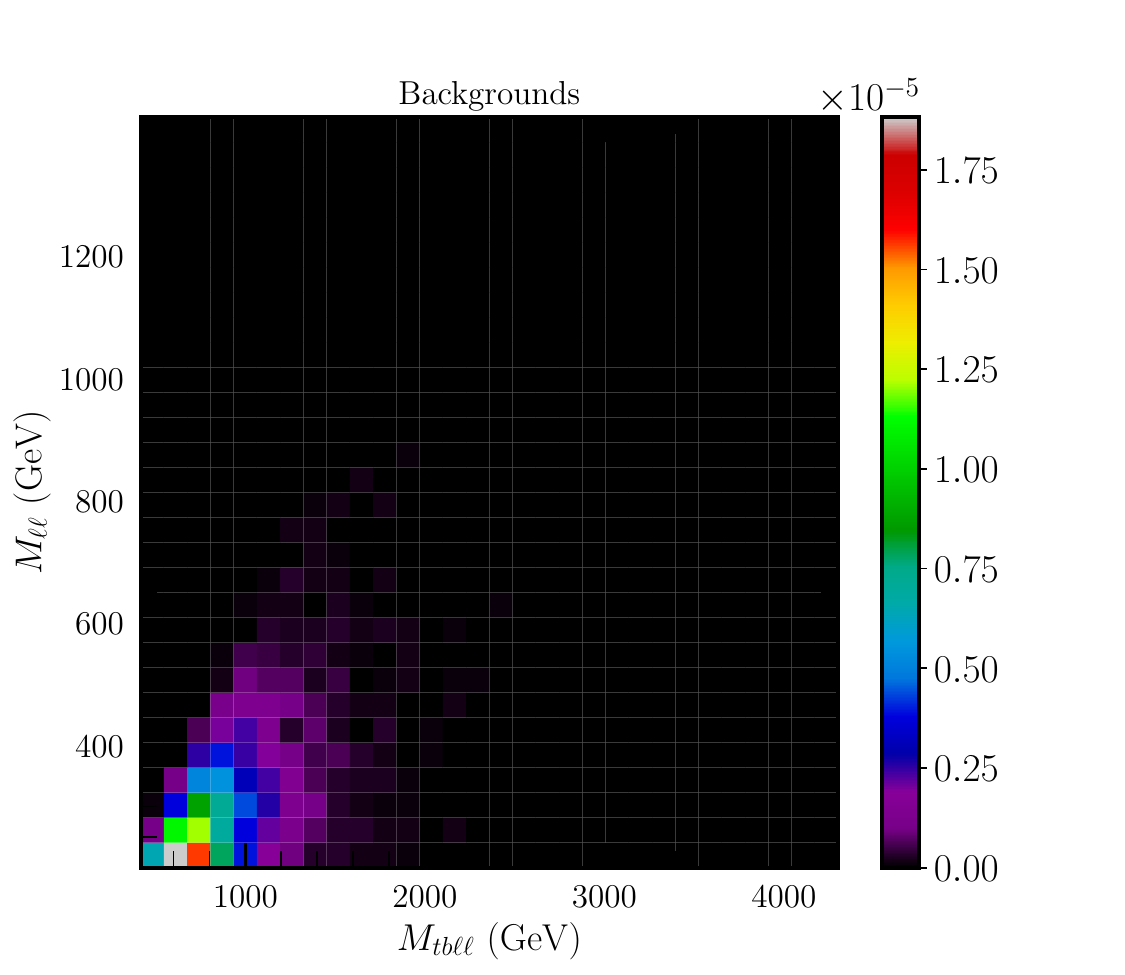}\hfill
  \includegraphics[width=0.45\linewidth,trim={7 0 60 0},clip]{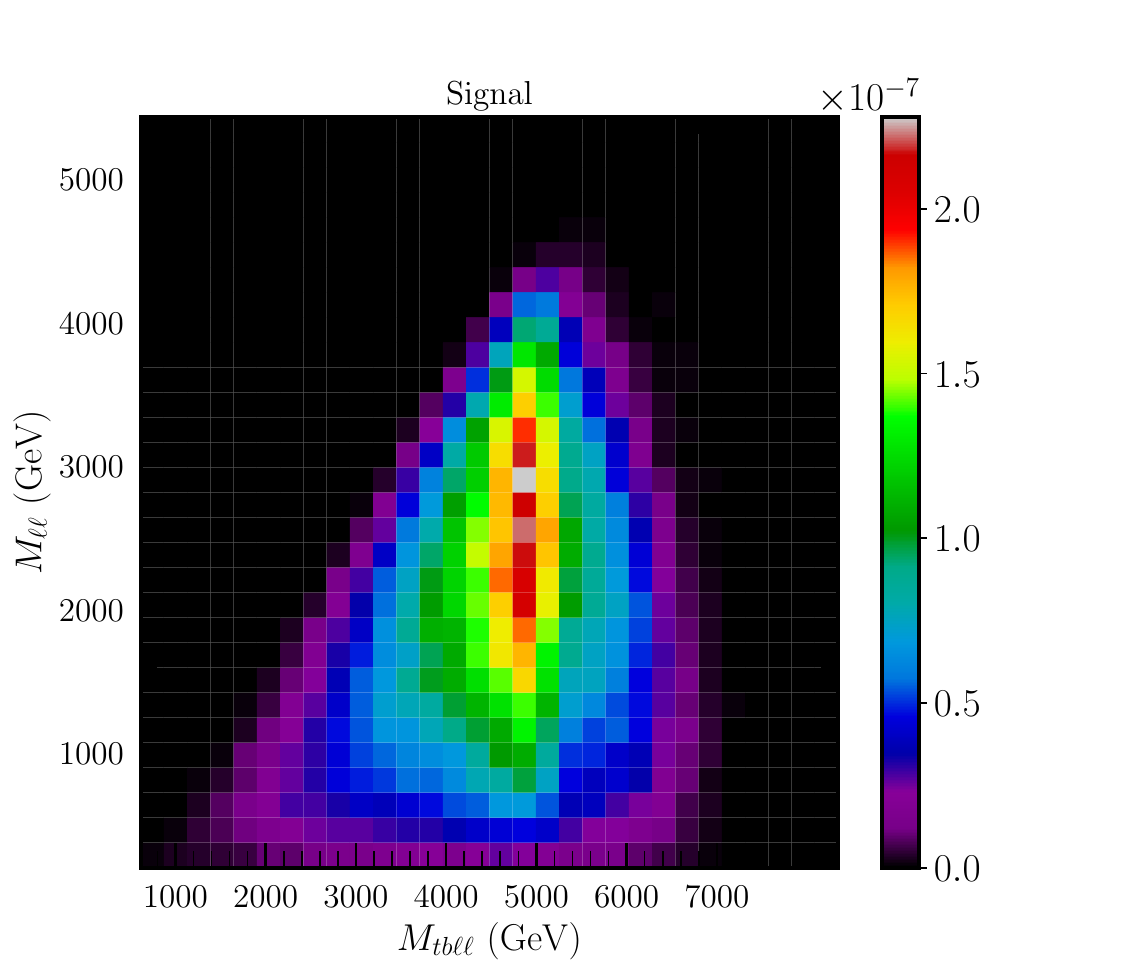}
  \caption{Normalised distributions in the invariant mass of the di-lepton system $M_{\ell\ell}$ (top left) and of the reconstructed $W_R$ boson $M_{\ell\ell tb}$ (top right) after the selection described in the text, both for the background (solid black) and the three representative signal scenarios BP$\mu$1 (blue), BP$\mu$2 (red) and BP$\mu$3 (orange). We additionally display the correlations between these two variables, both for the background (bottom left) and the BP$\mu$2 signal (bottom right). The colour code refers to a total number of entries in the maps, normalised to unity.}
  \label{fig:distributions:Mll:Mlltb}
\end{figure}

Background event generation and cross section calculations have been achieved with the same tool chain as the one described in section~\ref{sec:xsections}, with the exception of the NLO total cross sections relevant to single top production in association with a $Z$ boson or a $W$ boson that we extracted from Refs.~\cite{Campbell:2013yla} and \cite{Demartin:2016axk} respectively. Detector simulation relies on the SFS framework with a detector parametrisation matching the one that we used for the implementation of the CMS-EXO-20-002 search in \textsc{MadAnalysis}~5 (see Appendix~\ref{sec:MA5:CMS} for more details), with the exception of the $b$-tagging efficiency that we take equal to 70\%. Event reconstruction relies on the definition of two jet collections. The first one includes small-$R$ `AK04' jets clustered using the anti-$k_T$ algorithm and a jet radius parameter of $R=0.4$, while the second one comprises large-$R$ `CA15' jets clustered using the Cambridge-Aachen algorithm \cite{Dokshitzer:1997in, Wobisch:1998wt} with a radius parameter of $R=1.5$. Finally, boosted top quark candidates are reconstructed by means of \textsc{HepTopTagger}~\cite{Plehn:2010st,Kasieczka:2015jma}, as integrated within the jet substructure module of \textsc{MadAnalysis}~5~\cite{Araz:2023axv}.

%%%%%%%%%%%%%%%%%%%%%%%%%%%%%%%%%%%%%%%%
\subsection{Description of the analysis}
%%%%%%%%%%%%%%%%%%%%%%%%%%%%%%%%%%%%%%%%

\begin{table}
  \begin{center}
\setlength\tabcolsep{13pt}
    \begin{tabular}{l c c c}
 \diagbox{$M_{\ell\ell tb}$}{$M_{\ell\ell}$} &  $]400, \infty)$ & $]600, \infty)$ & $]800, \infty)$  \\
 \toprule
 $]1200, \infty)$ & SRa1 & SRb1 & SRc1 \\
 $]1400, \infty)$ & SRa2 & SRb2 & SRc2 \\
 $]1600, \infty)$ & SRa3 & SRb3 & SRc3 \\ 
 $]1800, \infty)$ & SRa4 & SRb4 & SRc4 \\
 $]2000, \infty)$ & SRa5 & SRb5 & SRc5 \\
 $]2500, \infty)$ & SRa6 & SRb6 & SRc6 \\
 $]3000, \infty)$ & SRa7 & SRb7 & SRc7 \\
    \end{tabular}
    \caption{Definitions of the SRs of our analysis in terms of bins in the invariant mass of the reconstructed $W_R$ boson $M_{\ell\ell tb}$ and the invariant mass of the lepton pair $M_{\ell\ell}$.}
    \label{tab:SR:definition}
  \end{center}
\end{table}

Events are selected if they contain exactly two isolated electrons or muons with $p_T > 25$~GeV for electrons and $p_T > 33$~GeV for muons, and $|\eta| < 2.4$ in both cases. Moreover, we remind that lepton isolation is encoded directly in our SFS detector parametrisation. The leading and sub-leading charged leptons are next enforced to satisfy $p_T > 60$~GeV and $p_T > 53$~GeV respectively, and we veto events featuring hadronically decaying tau-leptons with $p_T > 30$~GeV and $|\eta| < 2.5$. We then select events that contain at least one AK04 $b$-tagged jet with $p_T > 30$~GeV and $|\eta| < 2.5$, and at least one top-tagged CA15 jet with $p_T > 200$~GeV, $|\eta| < 2$, and an invariant mass $M_{\rm HTT} \in~]145, 210[$ GeV. Finally, we require that the invariant mass of the two charged leptons satisfies $M_{\ell\ell} > 200$~GeV. The resulting selection efficiency is $6.67\times 10^{-4}$ and $1.54 \times 10^{-3}$ for the $t+X$ and $t\bar{t}+X$ components of the background, respectively. On the other hand, for the signal processes the accumulated efficiency varies between $2.9\%$ and $5.4\%$ for the electron channel and between $14\%$ and $18\%$ for the muon channel, the higher efficiencies corresponding to scenarios with larger right-handed neutrino masses. Moreover, the higher prospects for the muon channel are directly related to the higher reconstructed efficiencies for high-$p_T$ muons than for corresponding electrons.

To improve signal significance we define several signal regions (SRs) targeting different bins in the invariant masses of the di-lepton system ($M_{\ell\ell}$) and of the reconstructed $W_R$ boson candidate ($M_{\ell\ell tb}$). To better understand this choice of variables, we display in Figure~\ref{fig:distributions:Mll:Mlltb} (top row) the associated normalised distributions for the background and a selection of three illustrative signal benchmark points (BP$\mu$1, BP$\mu$2 and BP$\mu$3), after imposing the selection $M_{\ell\ell} > 200$~GeV. The resonant contributions specific to the signal leads to very hard distributions exhibiting a very broad plateau (or peak), deep in the multi-TeV regime. This contrasts with the SM background, that exhibits a steeply falling spectrum with very few events expected for invariant masses larger than 1~TeV for the $M_{\ell\ell}$ spectrum or 2--3~TeV for the $M_{\ell\ell tb}$ one. The same information is displayed jointly through the correlations between these two variables in the bottom row of the figure, in which we focus on the background (bottom left) and the illustrative BP$\mu$2 scenario (bottom right). We subsequently define 21 SRs in each of the electron and muon channels as shown in Table~\ref{tab:SR:definition}. 

\begin{figure}
    \centering
    \includegraphics[width=0.85\linewidth]{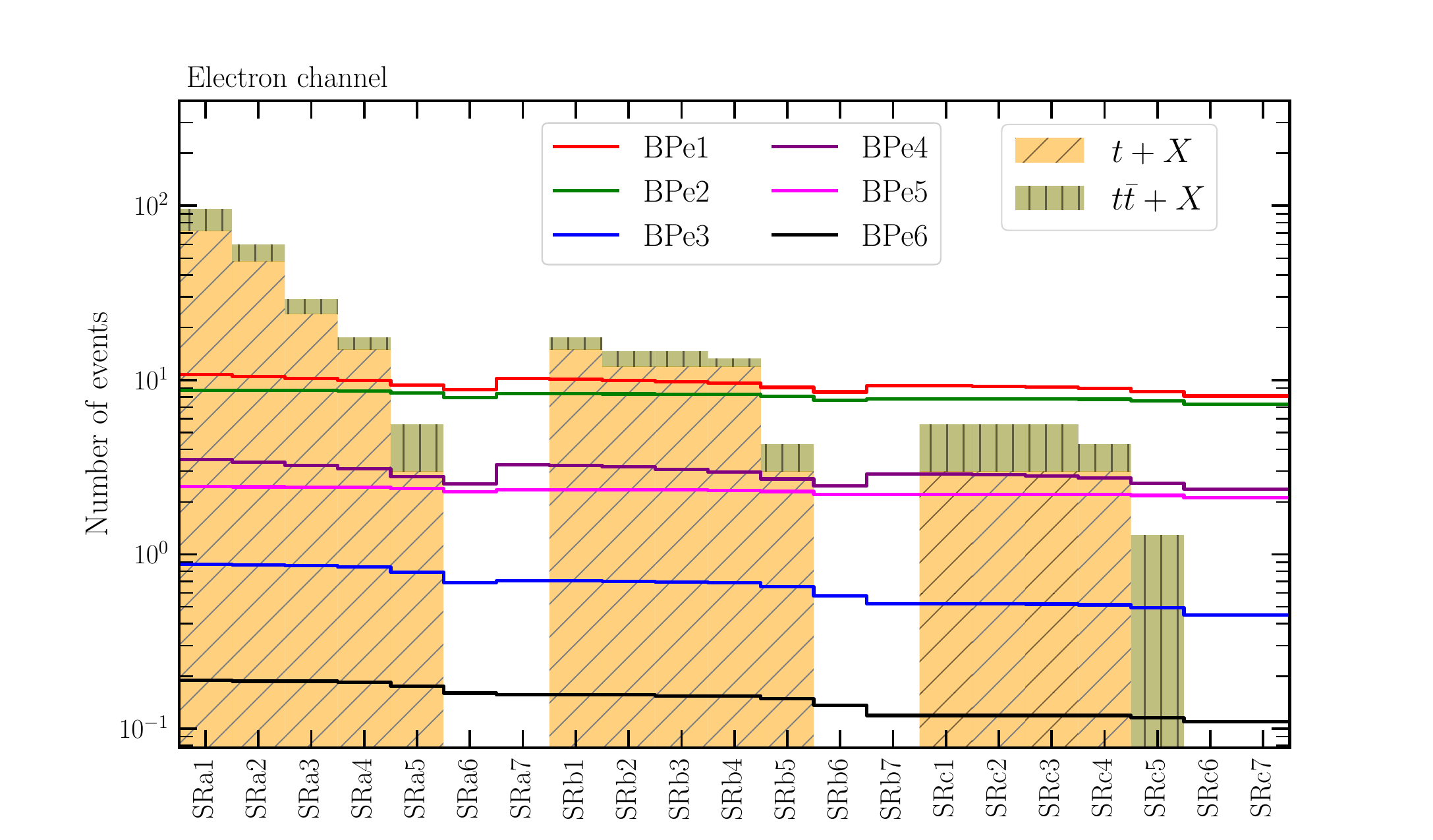}
    \includegraphics[width=0.85\linewidth]{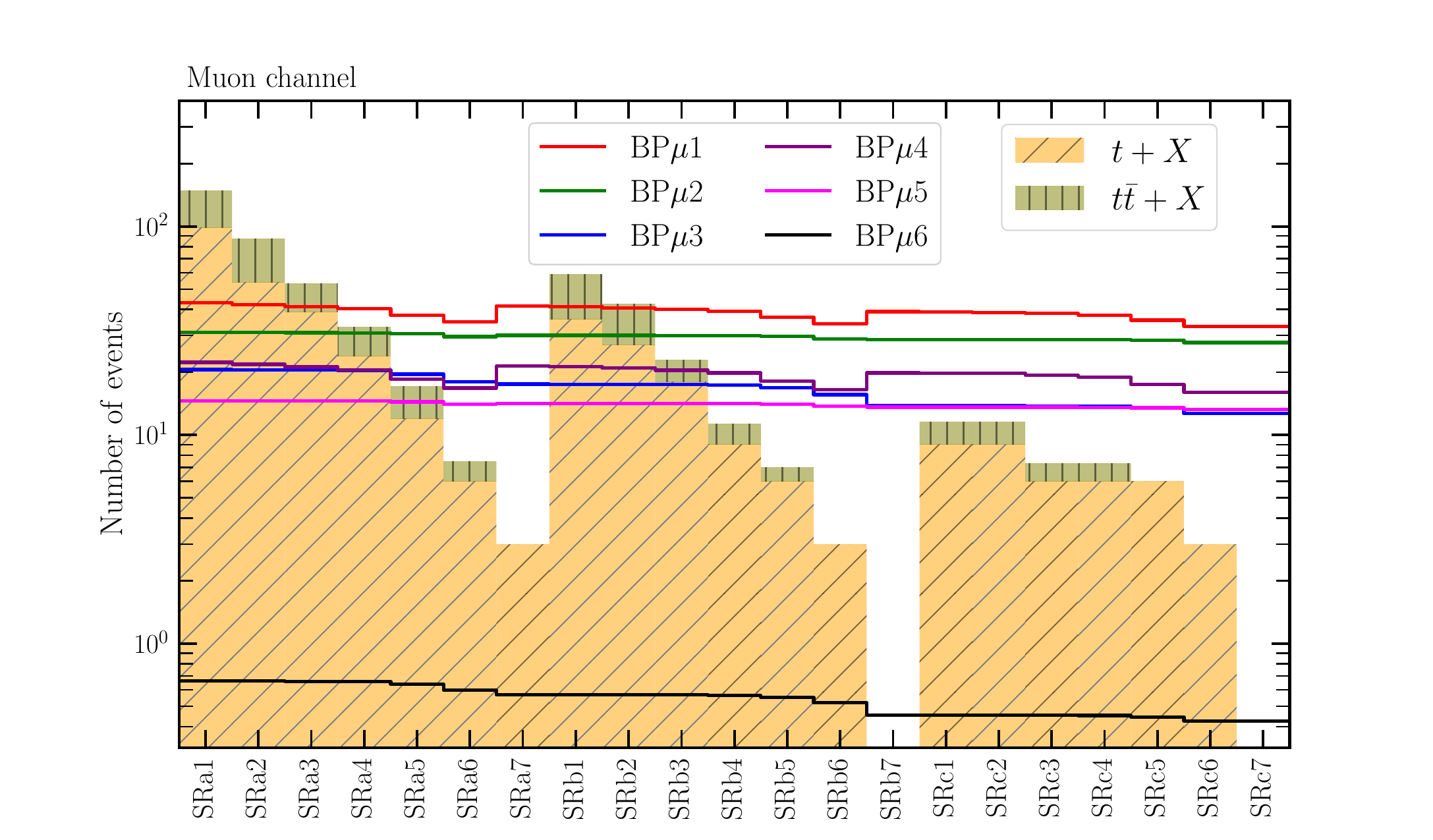}
    \caption{Number of background and signal events populating the different SRs in the electron (upper) and muon (lower) channels, for ${\cal L} = 3000$ fb$^{-1}$. Our predictions are normalised to NLO in QCD, and we collect the background contributions into a single top contribution (orange) and top-antitop contribution (kaki). Signal predictions are made for the 12 BPs considered (solid lines).}
    \label{fig:events:SR}
\end{figure}

%%%%%%%%%%%%%%%%%%%%%%%%%%%%%%%%%%%%%%%%
\subsection{Results}
%%%%%%%%%%%%%%%%%%%%%%%%%%%%%%%%%%%%%%%%

In Figure~\ref{fig:events:SR} we display the number of events populating the different signal regions of our analysis, both for the signal process in the case of the 12 benchmark points considered and the different contributions to the background. We collectively group the latter into a single top component (orange) and $t\bar t$ pair component (kaki), and we show results both for the electron channel (upper panel) and muon channel (lower panel). As already mentioned in section~\ref{sec:bench_bgd}, events are normalised to their total rates at NLO in the strong coupling. It is immediate to see that there is a high discovery potential for the signal, specifically in the muon channel. This originates from the fact that our analysis comprises some SRs that are essentially background-free, and that turn out to be the SRs dedicated to scenarios with heavy $W_R$ bosons.

In order to assess the sensitivity of the LHC to the signal process considered as a function of the integrated luminosity, we extract the ${\rm CL}_{\rm s}$ exclusion and associated $1\sigma$ and $2\sigma$ bands for the best expected signal region (SRb7), by means of the package \textsc{Pyhf}~\cite{pyhf_joss}. Whereas more aggressive estimates could be obtained by combining the different signal regions of the analysis, we refrain from doing so in order to get predictions as conservative as possible. Our goal is indeed to demonstrate that for scenarios not excluded by analyses of the $\ell\ell jj$ signature of $W_R$-boson production and decay, there is a potential gain in studying the corresponding third-generation signal $\ell\ell tb$. Our calculation relies on several assumptions. First, we conservatively assume that there is at least one background event regardless of the luminosity. Consequently, we always fix the number of background event to $n_b=1$ for signal regions in which our simulations yield $n_b<1$. Next, we assume that there is a systematic uncertainty of 20\% on the background yields. Finally, we estimate that the number of events to be observed is always equal to those predicted in the context of the SM. 

\begin{figure}
    \centering
    \includegraphics[width=0.95\linewidth,trim={105 0 90 0},clip]{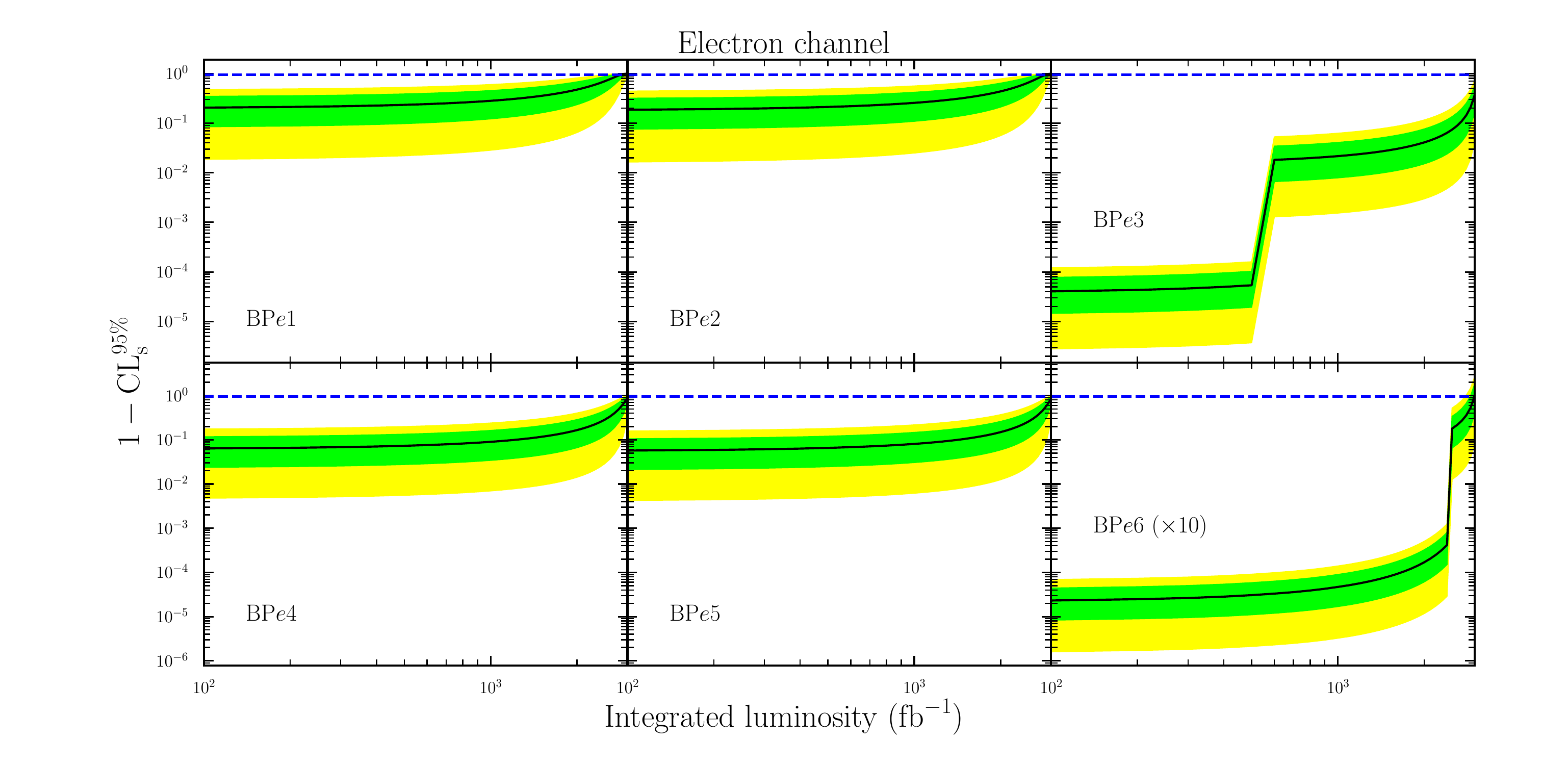}
    \includegraphics[width=0.95\linewidth,trim={105 0 95 0},clip]{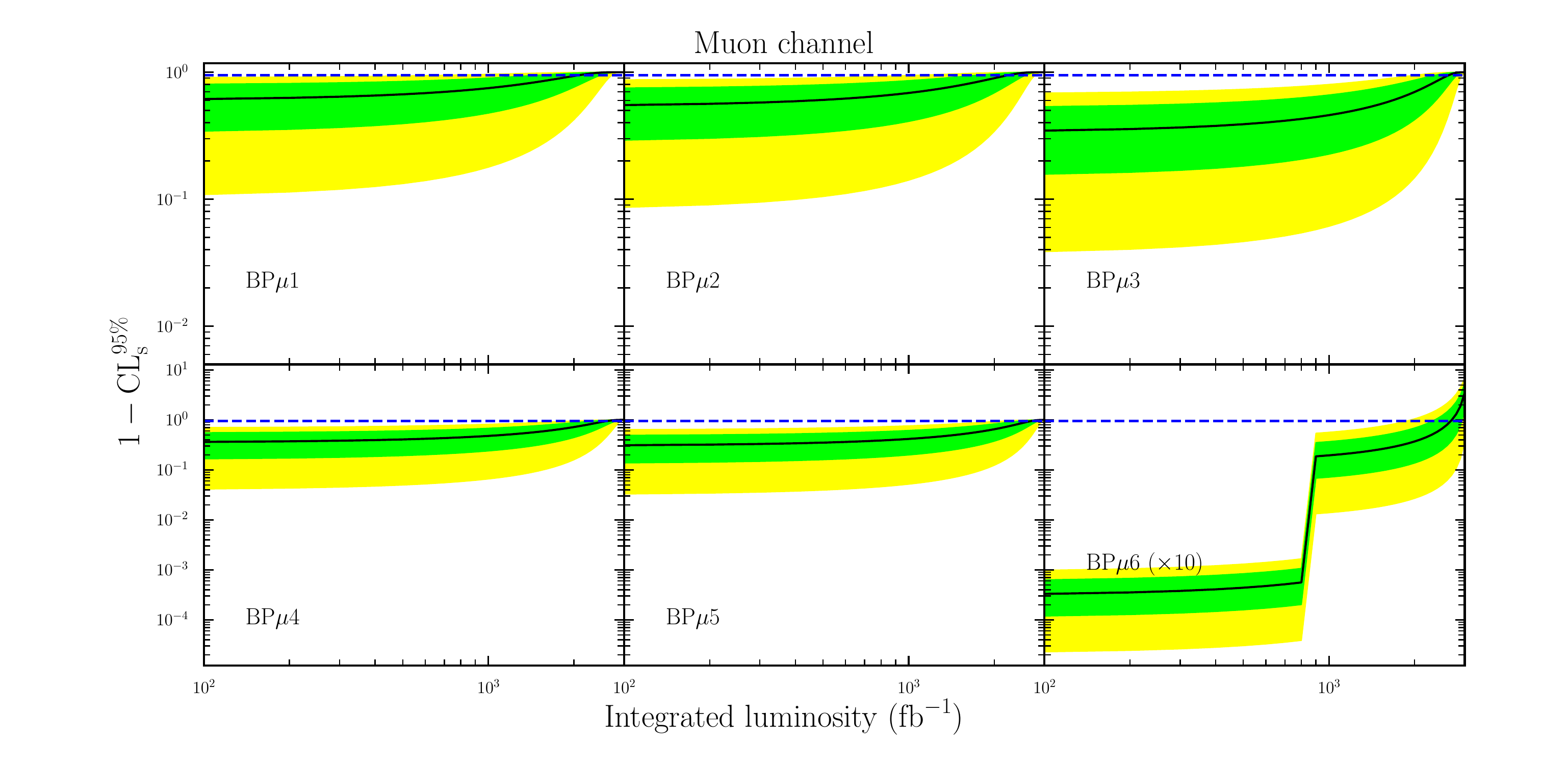}
    \caption{Evolution of the exclusion ${\rm CL}_{\rm s}$ as a function of the luminosity for the electron (upper) and muon (lower) channels, for each of the BPs considered. The central values are depicted through a solid black line, together with the associated $1\sigma$ (lime) and $2\sigma$ (yellow) bands. The horizontal blue dashed line corresponds to ${\rm CL}_{\rm s}=0.95$, above which the parameter point is excluded at $95\%$ confidence level.}
    \label{fig:CLs}
\end{figure}

The results are shown in Figure~\ref{fig:CLs} for the electron (upper panel) and muon (lower panel) channels. For the electron channel, it is clear that a large luminosity is required in order to probe the model. A few ab$^{-1}$ are indeed necessary to exploit the small signal cross sections (see Table~\ref{tab:BPs}). Consequently, only benchmarks for which the $W_R$ boson is not too heavy and in which the new physics spectrum is not too compressed could yield some sensitivity. We find that it is indeed the case, as shown by the predictions made in the upper panel of the figure. Only the two first scenarios, BP$e$1 and BP$e$2, have the potential to be excluded at 95\% confidence level, and this can only be achieved around the ultimate end of the high-luminosity LHC runs. In contrast, results in the muon channel are more promising, by virtue of the larger reconstruction efficiencies for high-$p_T$ muons than for electrons. More of our selected benchmarks are found reachable, and this for a smaller integrated luminosity (of about 1~ab$^{-1}$) than in the electron channel. All muonic scenarios with the exception of BP$\mu$6 are hence potentially observable, and it becomes clear that charged gauge boson masses ranging up to at least $W_R=5.5$~TeV could be probed during the high-luminosity phase of the LHC, provided that the mass splitting between the $W_R$ boson and the right-handed heavy neutrino $N_R$ is not too small. The $\ell\ell tb$ channel has thus a strong potential as a probe to left-right models, similar to the $\ell\ell jj$ channel (see Figure~\ref{fig:CLs:expected}).

\begin{figure}
    \centering
    \includegraphics[width=0.48\linewidth,trim={25 0 25 0},clip]{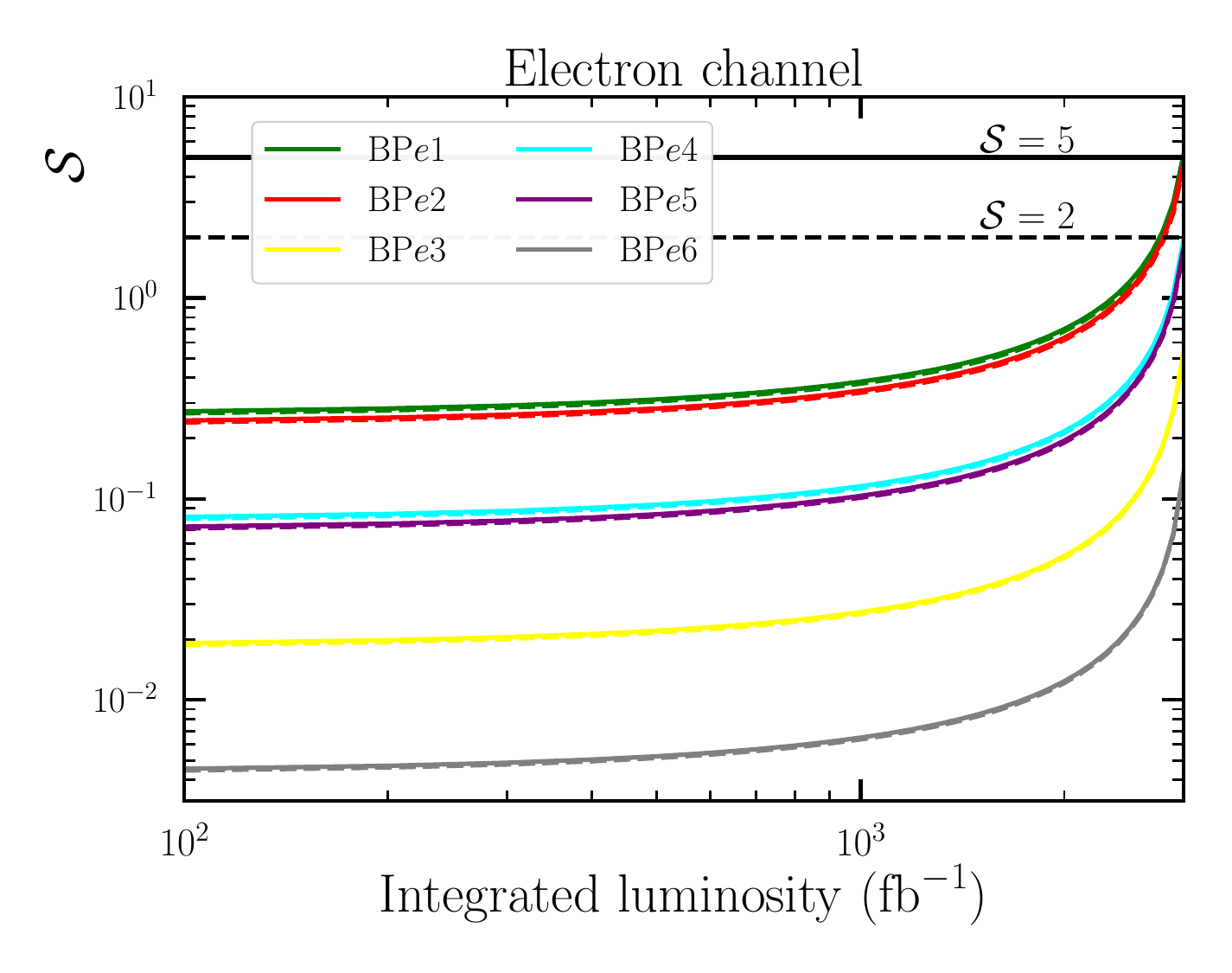}
    \hfill
    \includegraphics[width=0.48\linewidth,trim={25 0 25 0},clip]{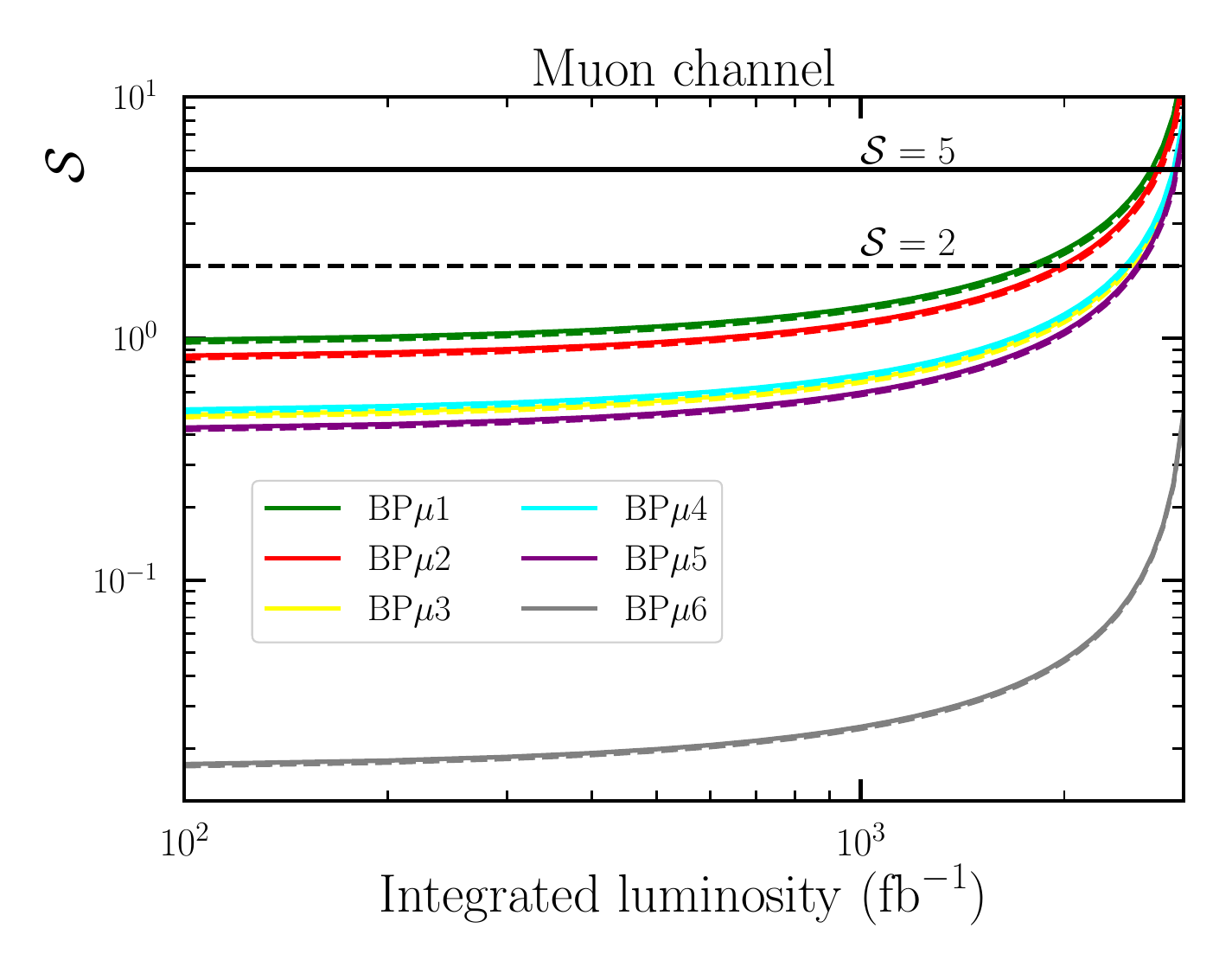}
    \caption{Signal significance ${\cal S}$ for the electron (left) and muon (right) channels. Results are shown for the scenarios BP$\ell$1 (green), BP$\ell$2 (red), BP$\ell$3 (yellow), BP$\ell$4 (cyan), BP$\ell$5 (purple) and BP$\ell$6 (gray) and for two choices of background uncertainties: the ideal case in which there is no error on the background (solid lines), and a more realistic situation embedding a level of systematics equal to $20\%$. The horizontal dashed and solid black lines correspond to the exclusion and discovery limits for which ${\cal S}=1$ and ${\cal S}=5$ respectively.}
    \label{fig:significance} 
\end{figure}

In Figure~\ref{fig:significance}, we illustrate these results in a complementary way through the calculation of the signal significance ${\cal S}$ as a function of the integrated luminosity, both for the electron (left panel) and muon (right panel) channel. Using the formulas introduced in Ref.~\cite{Cowan:2010js}, we make use of
\begin{equation}
  {\cal S}  =\sqrt{2}\left[(n_s+n_b)\log\left(\frac{(n_s+n_b)(n_b+\delta_b^2)}{n_b^2+(n_s+n_b)\delta_b^2}\right) - \frac{n_b^2}{\delta_b^2} \log\left(1+\frac{\delta_b^2 n_s}{n_b(n_b+\delta_b^2)}\right)\right]^{1/2},
\label{eq:SS:2}
\end{equation}
where $n_s$ and $n_b$ are the number of signal and background events, and where the systematic error on the background $\delta_b = x n_b$. We use the same error configuration as above, and we thus assume that $x=0.20$ and that $n_b$ is of at least 1 (dashed lines). For informative purposes, we additionally consider the ideal case of $x=0$ (solid lines). This consistently confirms that the two electron scenarios BP$e$1 and BP$e$2 can be reached at the high-luminosity LHC, and this for luminosities around $2.5$~ab$^{-1}$, and that the first five muonic BPs can be probed with luminosities of ${\cal O}(1)~{\rm ab}^{-1}$.

\begin{figure}
    \centering
    \includegraphics[width=0.32\linewidth,trim={5 0 60 0},clip]{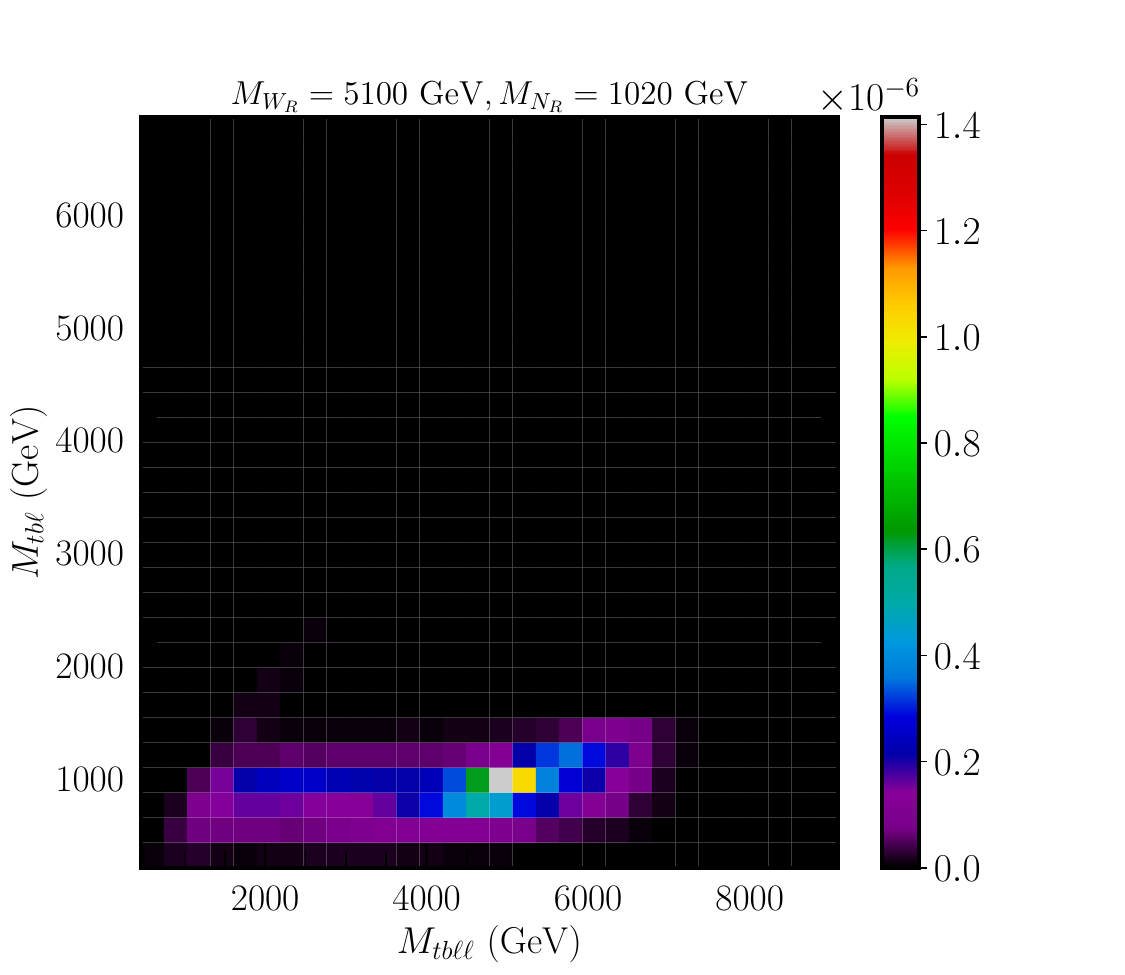}
    \includegraphics[width=0.32\linewidth,trim={5 0 60 0},clip]{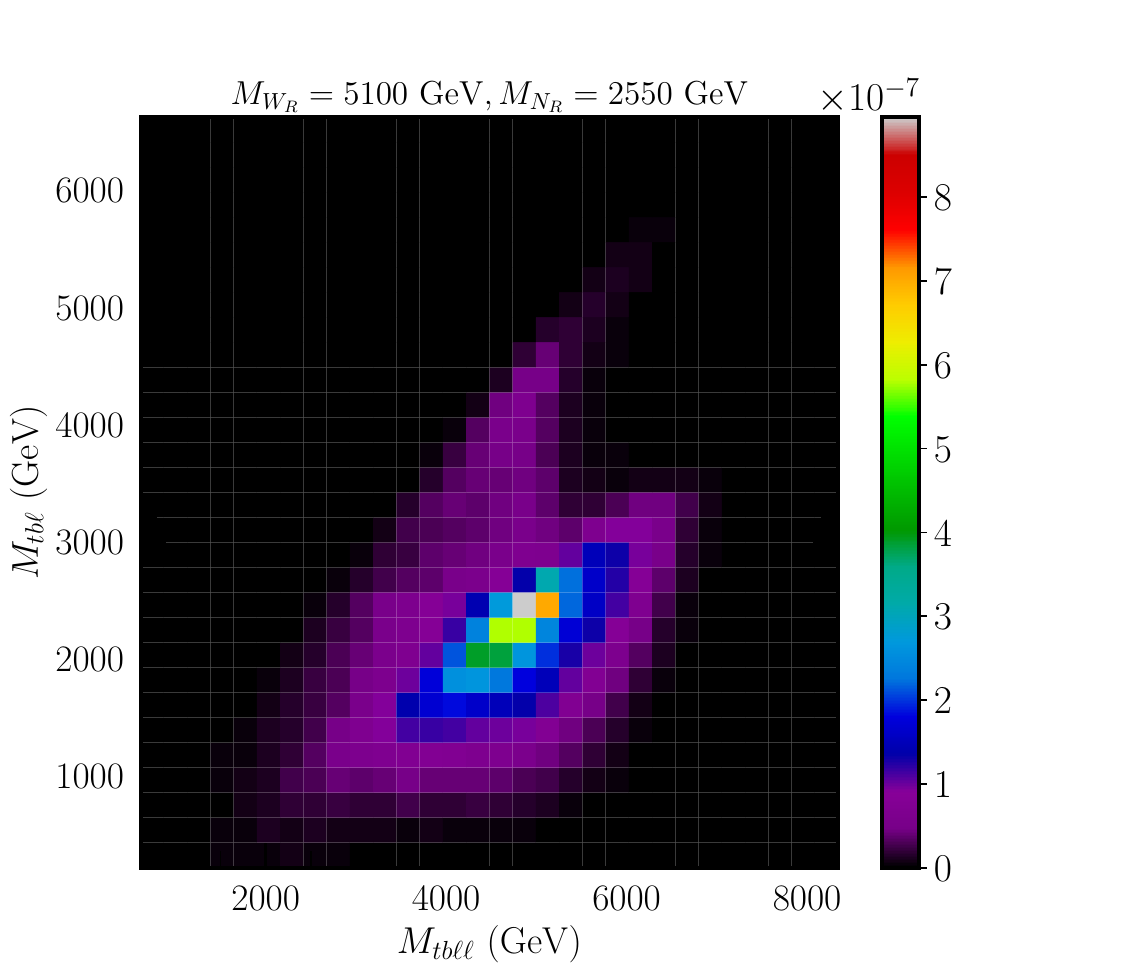}
    \includegraphics[width=0.32\linewidth,trim={5 0 60 0},clip]{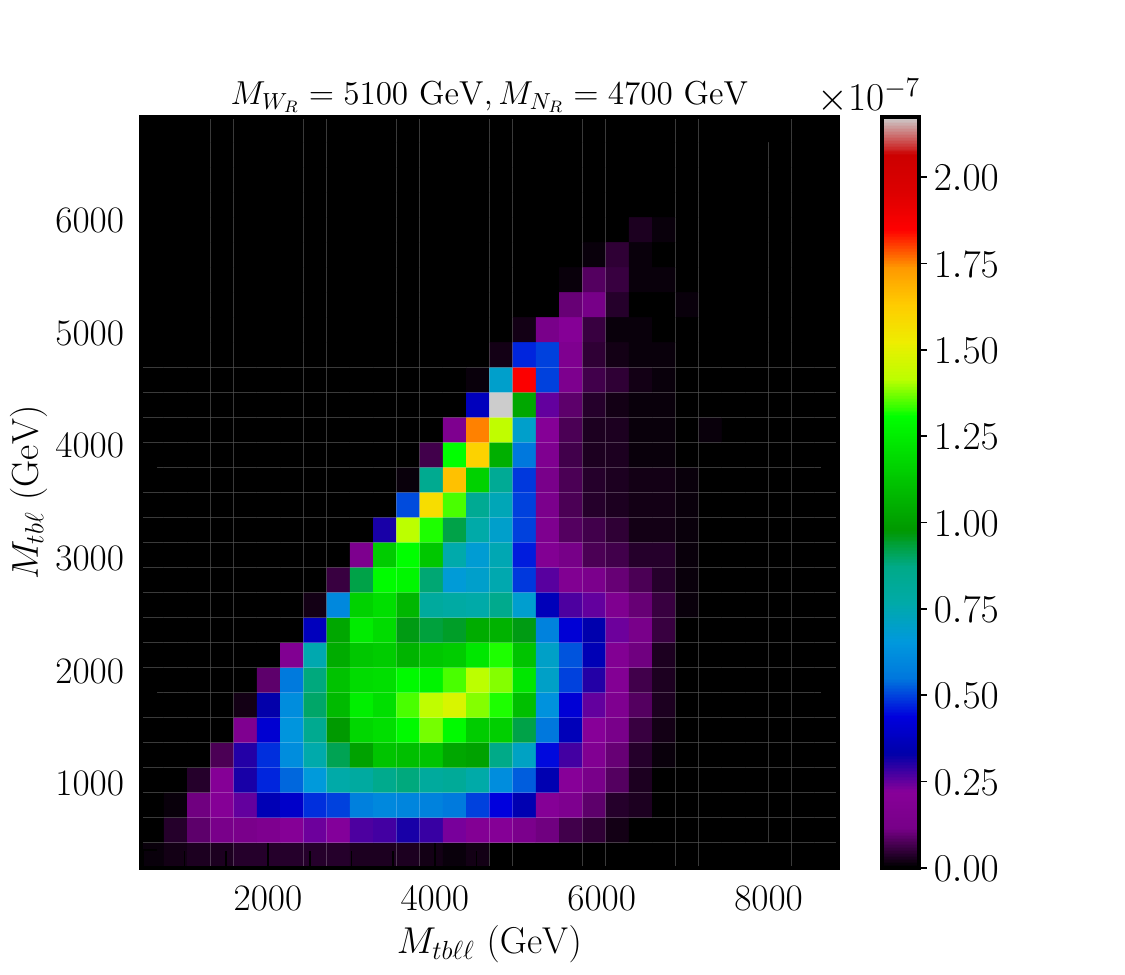}
    \caption{2D correlations between $M_{\ell\ell t b}$ and $M_{\ell tb}$ for the signal for BP$\mu$1 (left upper panel), BP$\mu$2 (right upper panel) and BP$\mu$3 (lower panel). The colour code corresponds to the number of entries normalised to unit integral.}
    \label{fig:correlation:Mtbll:Mtbl}
\end{figure}

We close this section by discussing briefly the capability of our search strategy to disentangle between the different BPs, {\it i.e.}\ by assessing the value of the right-handed neutrino mass. In our analysis we did not make use of the reconstructed $N_R$ mass, defined as the invariant mass of the $tb \ell$ system, for the reason that we wanted to optimise the signal-to-background ratio with minimal model assumptions. However, the determination of the $N_R$ mass can always be achieved post-discovery. In the production channel considered, the charged lepton that is emerging from the decay of the right-handed neutrino $N_R$, along with the top and bottom jets, can be tagged using a $\Delta R$ requirement: The $N_R$ candidate comprises the lepton with the minimal separation from the top-bottom system. As an example, we display in Figure~\ref{fig:correlation:Mtbll:Mtbl} correlation maps relating the invariant mass of the $W_R$ system ($M_{\ell\ell tb}$) and the one of the $N_R$ candidate ($M_{\ell tb}$), for the three most optimistic scenarios BP$\mu$1 (left panel), BP$\mu$2 (central panel) and BP$\mu$3 (right panel). These scenarios all feature the same $W_R$ boson mass, $M_{W_R} = 5.1$~TeV, but they differ with the chosen value of the right-handed neutrino mass so that different level of spectrum compression are considered. Predictions are presented for the signal region SRb7, that is essentially background-free and dedicated to scenarios with very large $W_R$-boson mass. This region corresponds to a mass window reasonably populated, which ensures good statistics for the signal ($n_{s} \approx 10$--$40$ events).  Our results therefore demonstrate that our simple analysis has an intrinsic ability to identify the underlying mLRSM scenario.
%%%%%%%%%%%%%%%%%%%%%%%%%%%%%%%%%
\section{Conclusions}
\label{sec:conclusions}
%%%%%%%%%%%%%%%%%%%%%%%%%%%%%%%%%

We have proposed a novel search channel for $SU(2)_R$ charged gauge bosons typical of left-right symmetric models, which exploits the fact that such bosons often decay, via right-handed neutrinos, into a final state made of a highly-boosted top quark, a $b$-jet and two charged leptons. This new avenue supplements traditional analyses that exploit the $\ell\ell jj$ signature, where $j$ represents a light jet. After discussing in detail the effects of existing searches on the viability of scenarios featuring a $W_R$ boson and a right-handed neutrino $N_R$ in a minimal left-right configuration, we have defined a few phenomenologically-viable scenarios that could be tested at future runs of the LHC. We have analysed the properties of $W_R$ and $N_R$ production and decay in these benchmark scenarios, and we have then built a new analysis strategy relying on the richness of the $\ell\ell tb$ final state. We have in particular made use of two kinematic variables to enhance the signal-to-background ratio, namely, the invariant masses of the di-lepton system ($M_{\ell\ell}$) and that of the $W_R$-boson candidate ($M_{\ell\ell tb}$). These variables have allowed for the definition of different SRs targeting different mLRSM spectra in terms of mass  scale and regime of compression, guaranteeing hence in a very  significant way a potential observation of the signal above the background. 

Our results demonstrated that analyses of the $\ell\ell tb$ channel could probe left-right models as well as analyses relying on the $\ell\ell jj$ channel. We have shown that, in the electron mode ($eetb$), scenarios featuring a $W_R$-boson with mass ranging up to $4.8$ TeV and a right-handed neutrino $N_R$ with mass lying between $M_{W_R}/5$ and $M_{W_R}/2$, have the potential to be probed during the high-luminosity phase of the LHC. Furthermore, in the muon channel ($\mu\mu tb$), $W_R$-boson masses ranging up to $5.5$ TeV can be reached for $N_R$ masses lying between $M_{W_R}/5$ and $M_{W_R}/2$, although $M_{N_R}$ can also sometimes be even larger. In addition, our approach is beneficial when it comes to profiling the underlining new physics scenario, as $N_R$ reconstruction is also possible. This in turn enables new physics characterisation through the exploitation of all information available from the rich final state inherent to $\ell\ell tb$ production via $W_R$ and $N_R$ exchanges, by virtue of the combination of $W_R$ and $N_R$ reconstruction.

The method that we proposed in this work can be improved further. First, better electron and muon identification efficiencies could increase fiducial signal rates, which would subsequently increase the sensitivity reach. This is especially crucial for the electron channel where lower reconstruction and identification efficiencies are currently in order. Second, one could use more sophisticated jet substructure techniques that will allow for the definition of the whole $N_R$ candidate as a single large $R$-jet with some specific substructure properties. Consequently, heavy-neutrino jets could be distinguished from more regular top-quark, $W$-boson, $Z$-boson or Higgs-boson jets. Such methods targeting jets with more than three prongs belong however to an uncharted territory, as far as we know, and they will need to be studied comprehensively. Third, we expect that machine-learning techniques could impact the signal significance very positively, and improve both large-$R$ jet identification and cut-flow optimisation. Finally, we remark that a simple search strategy, similar to the one that we proposed in this work, could also be used to probe lepton-flavour-violating or lepton-number-violating $W_R$-boson decays. All these items should however preferably be addressed in the context of an experimental search. We therefore urge the experimental collaborations at the LHC to carry out a search for $W_R$ bosons and right-handed neutrinos $N_R$ as we proposed.

\section*{Acknowledgements}
The work of AJ is supported by the Institute for Basic Science (IBS) under project code IBS-R018-D1. SM is supported in part through the NExT Institute and Science \& Technology Facilities Council (STFC) Consolidated Grant No.~ST/L000296/1. The work of MF has been partly supported by NSERC through the grant number SAP105354. BF acknowledges support from Grant ANR-21-CE31-0013, Project DMwithLLPatLHC, from the \emph{Agence Nationale de la Recherche} (ANR), France.

\appendix
%%%%%%%%%%%%%%%%%%%%%%%%%%%%%%%%%%%%%%%%%%%%%%%%%%%%%%%%%%%%%%%%%%%%%%%%%%%%%
\section{Reinterpretation of CMS searches for $W_R$ bosons with leptons and jets}
\label{sec:MA5:CMS}
%%%%%%%%%%%%%%%%%%%%%%%%%%%%%%%%%%%%%%%%%%%%%%%%%%%%%%%%%%%%%%%%%%%%%%%%%%%%%

In this section, we present details about predictions obtained with in-house implementations, in the \textsc{MadAnalysis}~5 framework~\cite{Conte:2012fm, Conte:2014zja, Conte:2018vmg}, of two CMS searches for $SU(2)_R$ gauge bosons and right-handed neutrinos in final states comprising two charged leptons (electrons or muons) and at least two jets. We consider LHC run~2 searches at $\sqrt{s} = 13$ TeV that focus on integrated luminosities of $35.9~{\rm fb}^{-1}$~\cite{CMS:2018agk} and $138~{\rm fb}^{-1}$~\cite{CMS:2021dzb}. We have validated these implementations by comparing cut-flows predicted with our tool chain to official results provided by the CMS collaboration for well-defined benchmark scenarios. The corresponding codes and the validation material is public, integrated in \textsc{MadAnalysis}~5, and can be found on the software dataverse~\cite{DVN/UMGIDL_2023,DVN/SGOK0J_2023}. In the following, we first discuss generalities common to the two implemented analyses in section \ref{app:common}, and we next detail the validation of the two implementations in sections \ref{app:CMS-EXO-17-011} and \ref{app:CMS-EXO-20-002}. 
 
%%%%%%%%%%%%%%%%%%%%%%%%%%%%%%%%%%%%%%%%%%%%%%%%%%%%%%%%%%%%
\subsection{Object definitions and detector modeling}
\label{app:common}
%%%%%%%%%%%%%%%%%%%%%%%%%%%%%%%%%%%%%%%%%%%%%%%%%%%%%%%%%%%%

The signal topology is such that the two produced final-state leptons carry a large fraction of the $SU(2)_R$ gauge boson energy. The two CMS analyses considered exploit this, and they consequently rely on triggers on leptons with very high transverse momentum to optimise signal efficiency. In CMS-EXO-17-011, the electron channel relies on a double-electron trigger that requires that the event final state features at least two electrons with a transverse momentum $p_T > 33$~GeV and associated to important deposits in the electromagnetic calorimeter. For the muon channel, a single-muon trigger is used instead, which requires the presence of at least one muon with $p_T > 50$~GeV. In its more recent CMS-EXO-20-002 search, the CMS collaboration improves electron triggering through a combination of three triggers. The event final state must exhibit the presence of at least either an isolated electron with $p_T > 27$~($32$)~GeV, an electron with $p_T > 115$ GeV, or a photon with $p_T > 175$ ($200$) GeV for the 2016~(2018) dataset.

As in all CMS searches, event reconstruction is performed by means of the particle-flow algorithm~\cite{CMS:2017yfk}. We briefly summarise below object identification requirements used in the two CMS searches considered, and we refer interested readers to Refs.~\cite{CMS:2018agk,CMS:2021dzb} for more details.\vspace{-.3cm}
\begin{itemize}
    \item Electron candidates are identified by associating charged-particle tracks from the primary vertex with energy deposit clusters in the electromagnetic calorimeter.\vspace{-.3cm}
    \item Muon candidates are identified by combining tracker and muon chamber information.\vspace{-.3cm}
    \item Charged hadrons reconstruction involves the matching of tracks with calorimeter cells, together with the absence of any associated activity in the muon chamber.\vspace{-.3cm}
    \item Neutral hadrons arises from the presence of clusters in both the electromagnetic and hadronic calorimeters and the absence of any associated charged particle track.\vspace{-.3cm}
    \item Jets are defined from the clustering of reconstructed particles by means of the anti-$k_T$ algorithm~\cite{Cacciari:2008gp, Cacciari:2011ma} with a jet radius $R=0.4$. Charged and neutral hadrons originating from pile-up interactions are removed through a dedicated pileup subtraction method~\cite{CMS:2017yfk} and residual average area-based corrections~\cite{Cacciari:2007fd}. % Agreement between reconstructed jet momentum and the simulated jet momentum is found to be of order $5\%$--$10\%$ across the whole transverse momentum spectrum and within the detector acceptance.
\end{itemize}

We model detector effects with the SFS module of \textsc{MadAnalysis}~5~\cite{Araz:2020lnp, Araz:2021akd}. Smearing and reconstruction efficiencies related to the different objects are tuned according to the CMS analyses considered, that we implement through approximate semi-analytical formulas. For energy and momentum smearing, we use the parametrisation included in the default CMS card shipped with \textsc{MadAnalysis}~5 (see also Refs.~\cite{CMS:2013lxn, CMS:2014pgm, CMS:2015xaf, CMS:2018rym}). We then implement the following identification efficiencies for jets (${\cal E}_{j}$), muons (${\cal E}_\mu$) and electrons (${\cal E}_e$):
\begin{equation}
\mathcal{E}_j = \begin{cases}
	 0.925 & \mathrm{if}\ |\eta| \leq 1.5, \\
	 0.875 & \mathrm{if}\,  1.5 < |\eta| \leq 2.5, \\ 
	 0.80  & \mathrm{if}\,  2.5 < |\eta|.
	\end{cases}
\end{equation}
\begin{equation}
  {\cal E}_e = \begin{cases}
		0.00 & \mathrm{for}\   p_{T} \leq 0.1~{\rm GeV}, \\
		0.73 & \mathrm{for}\,  |\eta| \leq 1.5 ~{\rm and}~ p_{T} \in ~]0.1, 1]~{\rm GeV}, \\ 
		0.95 & \mathrm{for}\,  |\eta| \leq 1.5 ~{\rm and}~ p_{T} \in ~]1, 100]~{\rm GeV}, \\
  		0.83 & \mathrm{for}\,  |\eta| \leq 1.5 ~{\rm and}~ p_{T} > 100~{\rm GeV}, \\
		0.50 & \mathrm{for}\,  1.5 < |\eta| \leq 2.5 ~{\rm and}~p_{T} \in ~]0.1, 1]~{\rm GeV}, \\
		0.83 & \mathrm{for}\,  1.5 < |\eta| \leq 2.5 ~{\rm and}~p_{T} \in ~]1, 100]~{\rm GeV}, \\
    	0.83 & \mathrm{for}\,  1.5 < |\eta| \leq 2.5 ~{\rm and}~p_{T} > 100~{\rm GeV}, \\
    	0.00 & \mathrm{for}\,  |\eta| > 2.5,
		\end{cases}
\end{equation}
\begin{equation}
{\cal E}_\mu =  \begin{cases}
		0.00 & \mathrm{for}\   p_T \leq 0.1~{\rm GeV}, \\
		0.75 & \mathrm{for}\,  |\eta| \leq 1.5 ~{\rm and}~ p_T \in ~]0.1, 1]~{\rm GeV}, \\ 
		0.99 & \mathrm{for}\,  |\eta| \leq 1.5 ~{\rm and}~ p_{T} \in ~]1, 100]~{\rm GeV}, \\
  		0.99 \times \exp\{0.5 - 5 \times 10^{-4} ~p_T\} & \mathrm{for}\,  |\eta| \leq 1.5 ~{\rm and}~ p_{T} > 100~{\rm GeV}, \\
		0.70 & \mathrm{for}\,  1.5 < |\eta| \leq 2.5 ~{\rm and}~p_{T} \in ~]0.1, 1]~{\rm GeV}, \\
		0.98 & \mathrm{for}\,  1.5 < |\eta| \leq 2.5 ~{\rm and}~p_{T} \in ~]1, 100]~{\rm GeV}, \\
    	0.98 \times \exp\{0.5 - 5 \times 10^{-4} ~p_T\} & \mathrm{for}\,  1.5 < |\eta| \leq 2.5 ~{\rm and}~p_{T} > 100~{\rm GeV}, \\
    	0.00 & \mathrm{for}\,  |\eta| > 2.5,
		\end{cases}
\end{equation}

%%%%%%%%%%%%%%%%%%%%%%%%%%%%%%%%
\subsection{CMS-EXO-17-011}
\label{app:CMS-EXO-17-011}
%%%%%%%%%%%%%%%%%%%%%%%%%%%%%%%%
\begin{table}
\setlength\tabcolsep{4pt}\renewcommand{\arraystretch}{1.1}
  \begin{center}
    \begin{tabular}{ll}
      Cut & Definition \\
      \toprule
      Initial & Initial number of events corresponding to $35.9~{\rm fb}^{-1}$\\
      $N_{\rm jets} \geq 2$ & At least two jets with $p_T > 25~{\rm GeV}$ and $|\eta| < 2.4$ \\ 
      $N_{\ell} \geq 2$ & At least two isolated charged leptons \\
      $p_T^j > 40$ GeV & The $p_T$ of the two leading jets should be larger than $40$ GeV \\ 
      $p_T^{\ell_1} > 60~{\rm GeV}$ & The $p_T$ of the leading lepton should be larger than 60~GeV  \\
      $p_T^{\ell_2} > 53~{\rm GeV}$ & The $p_T$ of the sub-leading lepton should be larger than 53~GeV\\
      $\Delta R_{\ell j} > 0.4$ & Leptons and jets should be separated by at least $\Delta R > 0.4$\\ 
      $M_{\ell \ell} > 200~{\rm GeV}$ & The di-lepton invariant mass should be larger than $ 200$ GeV\\
      $M_{\ell \ell jj} > 600~{\rm GeV}$ & The reconstructed $W_R$ invariant mass should be larger than $600$ GeV\\ 
    \end{tabular}
  \end{center}
  \caption{Definition of the different cuts used in the CMS--EXO--17--011 search~\cite{CMS:2018agk}.}
  \label{tab:cuts:CMS-EXO-17-011:definition}
\end{table}

The CMS-EXO-17-011 analysis is dedicated to events with at least two high-$p_T$ leptons and two jets. The two leptons, together with the two jets with the largest $p_T$, are considered to originate from the decay of a $W_R$ boson. The selection, that is described in Table~\ref{tab:cuts:CMS-EXO-17-011:definition}, first imposes that the leading and sub-leading leptons have transverse momentum $p_T > 60$ and $53$~GeV respectively, and that they both lie within the detector acceptance ($|\eta| < 2.4$, with electrons being rejected if $1.444 < |\eta| < 1.566$). Muon isolation is enforced by imposing that the charged-track activity in a cone of $\Delta R=0.3$ centred on the muon is of at most 10\% of the muon $p_T$, while electron isolation imposes that the same variable is smaller than $5$~GeV. On the other hand, the two jet candidates must each satisfy $p_T > 40$ GeV and $|\eta| < 2.4$. In addition, objects overlapping within $\Delta R > 0.4$ are removed. Signal region definition further constrains that the two leptons have the same flavour, that the invariant mass of the di-lepton system is above 200~GeV (to avoid contamination from the Drell-Yan background), and that the invariant mass of the reconstructed $W_R$ boson $M_{\ell \ell jj}$ is greater than 600~GeV. 

\begin{table}
\setlength\tabcolsep{3pt}\renewcommand{\arraystretch}{1.1}
\begin{center}
    \begin{tabular}{l cc cc cc}
      & \multicolumn{2}{c}{$M_{W_R}=2200$ GeV} & \multicolumn{2}{c}{$M_{W_R}=2800$ GeV} & \multicolumn{2}{c}{$M_{W_R}=3600$ GeV} \\
      Electron channel   & Events & $\varepsilon$ & Events & $\varepsilon$ & Events & $\varepsilon$ \\ 
      \toprule 
      Initial                                 & 1507.8 & $-$  & 366.2 & $-$  & 58.9 & $-$  \\
      $N_{\rm jets} \geq 2$                   & 1505.4 $ \pm $ 0.6 & 0.998 & 365.5 $ \pm $ 0.2 & 0.998  & 58.9 $ \pm $ 0.0 & 0.999 \\
      $N_{\ell} \geq 2$                       & 752.1 $ \pm $ 5.3 & 0.500 & 167.7 $ \pm $ 1.2 & 0.459 & 24.6 $ \pm $ 0.2 & 0.418 \\
      $p_{T}({\rm jets}) > 40~{\rm GeV}$      & 751.0 $ \pm $ 5.3 & 0.999 & 167.5 $ \pm $ 1.2 & 0.999 & 24.6 $ \pm $ 0.2 & 0.999 \\
      $p_{T}^{\ell_1} > 60~{\rm GeV}, p_{T}^{\ell_2} > 53~{\rm GeV}$& 735.1 $ \pm $ 5.3 & 0.979 & 164.8 $ \pm $ 1.2 & 0.984 & 24.4 $ \pm $ 0.2 & 0.992  \\
      $\Delta R_{\ell j} > 0.4$               & 735.1 $ \pm $ 5.3 & 1.000 & 164.8 $ \pm $ 1.2 & 1.000  & 24.4 $ \pm $ 0.2 & 1.000 \\
      $M_{\ell \ell} > 200~{\rm GeV}$         & 717.4 $ \pm $ 5.2 & 0.976 & 162.5 $ \pm $ 1.2 & 0.986  & 24.2 $ \pm $ 0.2 & 0.992 \\
      $M_{\ell \ell j j} > 600~{\rm GeV}$     & 717.0 $ \pm $ 5.2 & 0.999 & 162.5 $ \pm $ 1.2 & 1.000  & 24.2 $ \pm $ 0.2 & 1.000 \\[.4cm]
      
      Muon channel     & Events & $\varepsilon$ & Events & $\varepsilon$ & Events & $\varepsilon$ \\ 
      \toprule
      Initial                                 & 1507.8 & -  & 366.2 & -  & 58.9 & - \\
      $N_{\rm jets} \geq 2$                   & 1403.2 $ \pm $ 3.7 & 0.931 & 346.7 $ \pm $ 0.8 & 0.947 & 56.1 $ \pm $ 0.1 & 0.953 \\
      $N_{\ell} \geq 2$                       & 1213.0 $ \pm $ 5.4 & 0.864 & 300.8 $ \pm $ 1.3 & 0.867 & 46.3 $ \pm $ 0.2 & 0.824 \\
      $p_{T}({\rm jets}) > 40~{\rm GeV}$      & 1145.3 $ \pm $ 5.6 & 0.944 & 285.7 $ \pm $ 1.3 & 0.950 & 44.5 $ \pm $ 0.2 & 0.963 \\
      $p_{T}^{\ell_1} > 60~{\rm GeV}, p_{T}^{\ell_2} > 53~{\rm GeV}$& 1121.7 $ \pm $ 5.7 & 0.979 & 282.5 $ \pm $ 1.4 & 0.989 & 44.2 $ \pm $ 0.2 & 0.992 \\
      $\Delta R_{\ell j} > 0.4$               & 1121.7 $ \pm $ 5.7 & 1.000 & 282.5 $ \pm $ 1.4 & 1.000 & 44.2 $ \pm $ 0.2 & 1.000 \\
      $M_{\ell \ell} > 200~{\rm GeV}$         & 1103.7 $ \pm $ 5.7 & 0.984 & 279.5 $ \pm $ 1.4 & 0.989 & 43.9 $ \pm $ 0.2 & 0.994 \\
      $M_{\ell \ell j j} > 600~{\rm GeV}$     & 1103.7 $ \pm $ 5.7 & 1.000 & 279.5 $ \pm $ 1.4 & 1.000 & 43.9 $ \pm $ 0.2 & 1.000  \\
    \end{tabular}
    \caption{Cut-flow table relevant to the CMS-EXO-17-011 analysis, for the electron (upper) and muon (lower) channels. Three scenarios are considered, with $M_{W_R} = 2200$, 2800, and 3600~GeV and $M_{N_R} = M_{W_R} / 2$. Cut efficiencies are defined in Eq.~\eqref{eq:efficiency}.}
    \label{tab:cutflow:CMS-EXO-17-011}
  \end{center}
\end{table}

Cut-flow tables for selected benchmark scenarios are given in Table~\ref{tab:cutflow:CMS-EXO-17-011} for both the electron and the muon channels. We consider three $W_R$-boson masses of $2200$, $2800$ and $3600$~GeV, with the right-handed neutrino mass being fixed in each case to $M_{W_R} / 2$. We present the number of events surviving each cut, normalised to an integrated luminosity of $35.9$~fb$^{-1}$ and a production cross section evaluated at LO. Moreover, we define the efficiency $\varepsilon_i$ after the $i^\text{th}$ cut by
\begin{eqnarray}
\varepsilon_i = \frac{N_{i}}{N_{i-1}},
\label{eq:efficiency}
\end{eqnarray}
where $N_k$ corresponds to the number of events surviving the $k^\text{th}$ cut. In Table~\ref{tab:SR:ee:mm}, we compare the predictions obtained with \textsc{MadAnalysis}~5 for the benchmarks considered with the official results released by the CMS collaboration for different bins in the reconstructed $W_R$-boson mass. To better quantify the agreement we compute the quantity $\delta$ defined by
\begin{eqnarray}
\delta = 100 \times \bigg|1 - \frac{n_{\rm MA5}}{n_{\rm CMS}}\bigg|,
\label{eq:delta}
\end{eqnarray}
where $n_{\rm MA5}$~($n_{\rm CMS}$) refers to the number of events surviving all cuts and with $M_{\ell \ell jj} \in [M_{\ell \ell jj}^{\rm min}, M_{\ell \ell jj}^{\rm max}]$ as obtained with \textsc{MadAnalysis}~5 (reported by the CMS collaboration). We observe that our implementation leads to an agreement of about 2\%--20\%, and can thus be considered as validated.

\begin{table}
\setlength\tabcolsep{12pt}\renewcommand{\arraystretch}{1.1}
  \begin{center}
    \begin{tabular}{l cc l}
      $M_{W_R}$~($[M_{\ell \ell jj}^{\rm min}, M_{\ell \ell jj}^{\rm max}]$) & \textsc{MadAnalysis}~5 & CMS & $\delta~[\%]$ \\
      \toprule
      \multicolumn{4}{l}{Electron Channel} \\
      $2200$~([$1950, 2810$]) & $464.8 \pm 3.8$ & $474.0 \pm 3.7 \pm 44.7$ & $1.94$ \\
      $2800$~([$2530, 3840$]) & $117.2 \pm 0.9$ & $114.1 \pm 0.9 \pm 10.6$ & $2.71$ \\
      $3600$~([$3250, 5170$]) & $17.7 \pm 0.1$ & $19.2 \pm 0.2 \pm 1.8$ & $7.82$ \\[.2cm]
      \midrule 
      \multicolumn{4}{l}{Muon Channel} \\
      $2200$~([$1860, 2800$]) & $885.9 \pm 5.7$ & $744.0 \pm 4.7 \pm 47.5$ & $16.02$ \\ 
      $2800$~([$2430, 3930$]) & $211.2 \pm 1.4$ & $177.0 \pm 1.1 \pm 13.1$ & $16.19$ \\
      $3600$~([$3190, 5500$]) & $30.2 \pm 0.2$ & $29.2 \pm 0.2 \pm 2.6$ & $3.31$ \\
    \end{tabular}
  \end{center}
  \caption{Comparison between the number of events surviving all cuts as predicted by \textsc{MadAnalysis}~5, and those reported by the CMS collaboration. We consider three scenarios respectively featuring $M_{W_R} = 2200$, 2800 and 3600~GeV, and $M_{N_R} = M_{W_R} / 2$. For each scenario, a different bin in the invariant mass of the $\ell\ell jj$ system is considered.}
  \label{tab:SR:ee:mm}
\end{table}

%%%%%%%%%%%%%%%%%%%%%%%%%%%%%%%%
\subsection{CMS-EXO-20-002}
\label{app:CMS-EXO-20-002}
%%%%%%%%%%%%%%%%%%%%%%%%%%%%%%%%%

The CMS-EXO-17-011 analysis described in section~\ref{app:CMS-EXO-17-011} has been recently superseded by the CMS-EXO-20-002 search exploring similar signs of new physics, but in $138~{\rm fb}^{-1}$ of LHC run~2 data. This recent search for $W_R$-boson production and decay in an $\ell\ell jj$ system via a heavy neutrino includes two classes of search regions. A first one is dedicated to the resolved regime where (at least) four well-separated final-state objects are identified, while a second one focuses on a boosted situation in which (at least) only two well-separated objects are identified. We only consider the former, as it is sufficient for the present study. Report on the recasting of the boosted-regime analysis is left for future work. 

The resolved analysis is similar the CMS-EXO-17-011 search, although it embeds a few differences. Events are selected provided that their final state includes exactly two isolated leptons, and at least two small-radius jets (with $R=0.4$) with $p_T > 40$~GeV and $|\eta| < 0.4$. Among all jets, those with the highest transverse momenta are assumed to originate from the $W_R$-boson decay, together with the leading and sub-leading charged leptons that are required to satisfy $p_T > 60$~GeV and 53~GeV respectively, and to be within $|\eta| < 2.4$ (and not to satisfy $1.44 < |\eta| < 1.57$ for electrons). Lepton isolation is enforced through a dedicated variable 
\begin{eqnarray}
    I_\ell \equiv \sum_{i\in \mathrm{tracks}} p_T^i,
\end{eqnarray}
where the sum runs over all tracks separated by $\Delta R < 0.3$ from the lepton direction. Lepton definition then requires that $I_\mu < 0.1 p_T^\mu$, and that $I_e < 5$~GeV. Furthermore, all final-state object candidates are imposed not to overlap, and to satisfy $\Delta R > 0.4$. In addition, a good background rejection is guaranteed by a first selection on the di-lepton invariant mass, $M_{\ell\ell} > 400$~GeV, and by a second selection on the invariant mass of the reconstructed $W_R$-boson candidate (\textit{i.e.}\ the invariant mass of the $\ell\ell jj$ system), $M_{\ell\ell jj} > 800$~GeV.

\begin{table}
\setlength\tabcolsep{7pt}\renewcommand{\arraystretch}{1.15}
  \begin{center}
    \begin{tabular}{l cc ccc}
      & \multicolumn{2}{c}{CMS} & \multicolumn{3}{c}{\textsc{MadAnalysis}} \\
      \toprule 
      $\{M_{W_R},M_{N_R}\} = (3000, 1400)~{\rm GeV}$ & Events & $\varepsilon$ & Events & $\varepsilon$ & $\delta$ [\%]\\ 
      Initial                                 & 1175.4 & -  & 1174.4 & - & - \\
      $N_{\ell} \geq 2~{\rm with}~p_{T}^{\ell} > 60~(53)~{\rm GeV}$& 379.5  & 0.323 & 363.5 $ \pm $ 3.0 & 0.309 & 4.1 \\
      $\geq 2~{\rm AK4~jets~with}~p_T > 40~{\rm GeV}$& 363.1  & 0.957 & 363.0 $ \pm $ 3.0 & 0.999 & 4.4 \\
      $\Delta R > 0.4~{\rm between~all~pairs~of~objects}$& 355.4  & 0.979 & 363.0 $ \pm $ 3.0 & 1.000 & 2.2 \\
      $m_{\ell\ell} > 200~{\rm GeV}$          & 335.6  & 0.944 & 358.3 $ \pm $ 3.0 & 0.987 & 4.5 \\
      $m_{\ell\ell jj} > 800~{\rm GeV}$       & 335.6  & 1.000 & 357.1 $ \pm $ 3.0 & 0.997 & 0.3 \\
      $m_{\ell\ell} > 400~{\rm GeV}$          & 324.1  & 0.966 & 339.5 $ \pm $ 2.9 & 0.951 & 1.6 \\[.2cm]
      $\{M_{W_R},M_{N_R}\} = (4000, 2000)~{\rm GeV}$ & Events & $\varepsilon$ & Events & $\varepsilon$ & $\delta$ [\%]\\ 
      \toprule
      Initial                                 & 115.2 & -  & 115.2 & - & - \\
      $N_{\ell} \geq 2~{\rm with}~p_{T}^{\ell} > 60~(53)~{\rm GeV}$& 37.2  & 0.322 & 32.9 $ \pm $ 0.3 & 0.285 & 11.5 \\
      $\geq 2~{\rm AK4~jets~with}~p_T > 40~{\rm GeV}$& 36.0  & 0.969 & 32.7 $ \pm $ 0.3 & 0.996 & 2.8 \\
      $\Delta R > 0.4~{\rm between~all~pairs~of~objects}$& 35.1  & 0.974 & 32.7 $ \pm $ 0.3 & 1.000 & 2.6 \\
      $m_{\ell\ell} > 200~{\rm GeV}$          & 33.2  & 0.947 & 32.5 $ \pm $ 0.3 & 0.993 & 4.8 \\
      $m_{\ell\ell jj} > 800~{\rm GeV}$       & 33.2  & 1.000 & 32.5 $ \pm $ 0.3 & 0.999 & 0.1 \\
      $m_{\ell\ell} > 400~{\rm GeV}$& 32.5  & 0.979 & 31.4 $ \pm $ 0.3 & 0.966 & 1.3 \\[.2cm]
      $\{M_{W_R},M_{N_R}\} = (5000, 3000)~{\rm GeV}$ & Events & $\varepsilon$ & Events & $\varepsilon$ & $\delta$ [\%]\\ 
      \toprule
      Initial                                 & 11.8 & -  & 11.8 & - & - \\
      $N_{\ell} \geq 2~{\rm with}~p_{T}^{\ell} > 60~(53)~{\rm GeV}$& 3.8 & 0.317 & 3.1 $ \pm $ 0.0 & 0.264 & 16.9 \\
      $\geq 2~{\rm AK4~jets~with}~p_T > 40~{\rm GeV}$ & 3.7  & 0.987 & 3.1 $ \pm $ 0.0 & 0.999 & 1.3 \\
      $\Delta R > 0.4~{\rm between~all~pairs~of~objects}$& 3.6  & 0.968 & 3.1 $ \pm $ 0.0 & 1.000 & 3.4 \\
      $m_{\ell\ell} > 200~{\rm GeV}$          & 3.4 & 0.947 & 3.1 $ \pm $ 0.0 & 0.997 & 5.3 \\
      $m_{\ell\ell jj} > 800~{\rm GeV}$       & 3.4  & 1.000 & 3.1 $ \pm $ 0.0 & 1.000 & 0.0 \\
      $m_{\ell\ell} > 400~{\rm GeV}$& 3.4  & 0.988 & 3.0 $ \pm $ 0.0 & 0.974 & 1.4 \\
    \end{tabular}
    \caption{Cut-flow tables as predicted with \textsc{MadAnalysis}~5 and as reported by the CMS collaboration, for the analysis of the $eejj$ final state. Three scenarios are considered, with $\{M_{W_R}, M_{N_R}\} = (3000, 1400)$, $(4000, 2000)$ and $(5000, 3000)$~GeV, for which we provide number of events normalised to $138~{\rm fb}^{-1}$ and cut-by-cut efficiencies ($\varepsilon$) as defined in Eq.~\eqref{eq:efficiency}. We also display the level of agreement between the CMS and \textsc{MadAnalysis}~5 results, as quantified by the $\delta$ variabble introduced in Eq.~\eqref{eq:delta}.}
    \label{tab:cutflow:ee:CMS-EXO-20-002}
  \end{center} 
\end{table}

\begin{table}
\setlength\tabcolsep{7pt}\renewcommand{\arraystretch}{1.15}
  \begin{center}
    \begin{tabular}{l cc ccc}
      & \multicolumn{2}{c}{CMS} & \multicolumn{3}{c}{{\textsc{MadAnalysis}}} \\ 
      \toprule
     $\{M_{W_R},M_{N_R}\} = (3000, 1400)~{\rm GeV}$ & Events & $\varepsilon$ & Events & $\varepsilon$ & $\delta$ [\%]\\ 
      Initial                                 & 586.6 & -  & 587.2 & - & - \\
      $N_{\ell} \geq 2~{\rm with}~p_{T}^{\ell} > 60~(53)~{\rm GeV}$& 494.5 & 0.843 & 498.6 $ \pm $ 1.9 & 0.849 & 0.7 \\
      $\geq 2~{\rm AK4~jets~with}~p_T > 40~{\rm GeV}$& 473.2  & 0.957 & 450.6 $ \pm $ 2.2 & 0.904 & 5.6 \\
      $\Delta R > 0.4~{\rm between~all~pairs~of~objects}$& 443.1  & 0.936 & 450.6 $ \pm $ 2.2 & 1.000 & 6.8 \\
      $m_{\ell\ell} > 200~{\rm GeV}$          & 420.8  & 0.950 & 447.3 $ \pm $ 2.2 & 0.993 & 4.5 \\
      $m_{\ell\ell jj} > 800~{\rm GeV}$       & 420.7  & 1.000 & 447.2 $ \pm $ 2.2 & 1.000 & 0.0 \\
      $m_{\ell\ell} > 400~{\rm GeV}$ & 407.3  & 0.968 & 431.6 $ \pm $ 2.2 & 0.965 & 0.3 \\[.2cm]
      $\{M_{W_R},M_{N_R}\} = (4000, 2000)~{\rm GeV}$     & Events & $\varepsilon$ & Events & $\varepsilon$ & $\delta$ [\%]\\ \toprule
      Initial                                 & 57.6 & -  & 57.6 & - & - \\
      $N_{\ell} \geq 2~{\rm with}~p_{T}^{\ell} > 60~(53)~{\rm GeV}$& 49.8 & 0.864 & 45.2 $ \pm $ 0.2 & 0.786 & 9.1 \\
      $\geq 2~{\rm AK4~jets~with}~p_T > 40~{\rm GeV}$& 48.4 & 0.971 & 41.8 $ \pm $ 0.2 & 0.925 & 4.8 \\
      $\Delta R > 0.4~{\rm between~all~pairs~of~objects}$& 45.4  & 0.940 & 41.8 $ \pm $ 0.2 & 1.000 & 6.4 \\
      $m_{\ell\ell} > 200~{\rm GeV}$          & 43.5  & 0.957 & 41.7 $ \pm $ 0.2 & 0.997 & 4.2 \\
      $m_{\ell\ell jj} > 800~{\rm GeV}$       & 43.5  & 1.000 & 41.7 $ \pm $ 0.2 & 1.000 & 0.0 \\
      $m_{\ell\ell} > 400~{\rm GeV}$ & 42.8  & 0.985 & 40.7 $ \pm $ 0.2 & 0.977 & 0.8 \\[.2cm]   
      $\{M_{W_R},M_{N_R}\} = (5000, 3000)~{\rm GeV}$      & Events & $\varepsilon$ & Events & $\varepsilon$ & $\delta$ [\%]\\ \toprule
      Initial                                 & 5.9 & -  & 5.9 & - & - \\
      $N_{\ell} \geq 2~{\rm with}~p_{T}^{\ell} > 60~(53)~{\rm GeV}$& 5.1  & 0.858 & 4.4 $ \pm $ 0.0 & 0.749 & 12.7 \\
      $\geq 2~{\rm AK4~jets~with}~p_T > 40~{\rm GeV}$& 5.0  & 0.984 & 4.1 $ \pm $ 0.0 & 0.923 & 6.2 \\
      $\Delta R > 0.4~{\rm between~all~pairs~of~objects}$& 4.7  & 0.940 & 4.1 $ \pm $ 0.0 & 1.000 & 6.4 \\
      $m_{\ell\ell} > 200~{\rm GeV}$          & 4.5  & 0.960 & 4.1 $ \pm $ 0.0 & 0.997 & 3.9 \\
      $m_{\ell\ell jj} > 800~{\rm GeV}$       & 4.5  & 1.000 & 4.1 $ \pm $ 0.0 & 1.000 & 0.0 \\
      $m_{\ell\ell} > 400~{\rm GeV}$          & 4.5  & 0.991 & 4.0 $ \pm $ 0.0 & 0.982 & 0.9 \\
    \end{tabular}
    \caption{Same as in Table~\ref{tab:cutflow:ee:CMS-EXO-20-002} but for the $\mu\mu jj$ channel.}
    \label{tab:cutflow:mm:CMS-EXO-20-002}
  \end{center}
\end{table}

A comparison between the results obtained with our implementation of the CMS-EXO-20-002 analysis in \textsc{MadAnalysis}~5 and the official results as released by the CMS collaboration are shown in Tables~\ref{tab:cutflow:ee:CMS-EXO-20-002} (for the electron channel) and \ref{tab:cutflow:mm:CMS-EXO-20-002} (for the muon channel). In these tables, we consider three scenarios defined by $\{M_{W_R}, M_{N_R}\} = (3000,1400)$, $(4000,2000)$ and $(5000,3000)$~GeV. To quantify the agreement between our predictions with \textsc{MadAnalysis}~5 and the official results provided by the CMS collaboration,  we make use of the $\delta$ variable introduced in Eq.~\eqref{eq:delta}. An amazingly good agreement between the two is found, which validates our implementation.

\bibliographystyle{JHEP}
\bibliography{biblio.bib}

%%%%%%%%%%%%%%%%%%%%%%%%%%%%
\end{document}